\documentclass[printer]{aa}
\usepackage[varg]{txfonts}
\usepackage{natbib} 
\usepackage[pdftex]{graphicx} 
\usepackage{booktabs} 
\usepackage{subfigure}
\usepackage{gensymb}
\usepackage{amstext}
\usepackage{slashbox}

\bibpunct{(}{)}{;}{a}{}{,} 
\usepackage{color}
\begin{document} 

\title{Overview and stellar statistics of the expected Gaia Catalogue using the Gaia Object Generator}

\author{X. Luri    \inst{1}
\and M. Palmer     \inst{1}
\and F. Arenou     \inst{2}
\and E. Masana     \inst{1}
\and J. de Bruijne \inst{3} 
\and E. Antiche    \inst{1}
\and C. Babusiaux  \inst{2}
\and R. Borrachero \inst{1}
\and P. Sartoretti \inst{2}
\and F. Julbe      \inst{1}
\and Y. Isasi      \inst{1}
\and O. Martinez   \inst{1}
\and A.C. Robin    \inst{4}
\and C. Reyl\'{e}  \inst{4}
\and C. Jordi      \inst{1}
\and J.M. Carrasco \inst{1}}

\institute{Dept. d'Astronomia i Meteorologia, Institut de Ci\`{e}ncies del Cosmos, Universitat de Barcelona (IEEC-UB), Mart\'{i} Franqu\`{e}s 1, E08028 Barcelona, Spain.\label{1}
\and GEPI, Observatoire de Paris, CNRS, Universit\'{e} Paris Diderot, 5 place Jules Janssen, 92190, Meudon, France \label{2}
\and Scientific Support Office of the Directorate of Science and Robotic Exploration of the European Space Agency, European Space Research and Technology Centre, Keplerlaan 1, 2201 AZ, Noordwijk, The Netherlands \label{3}
\and Institut UTINAM, CNRS UMR 6213, Observatoire des Sciences de l'Univers THETA Franche-Comt\'{e}-Bourgogne, Universit\'e de Franche Comt\'{e}, Observatoire de Besan\c{c}on, BP 1615, 25010 Besan\c{c}on Cedex, France \label{4}}


\abstract{}
{An effort has been made to simulate the expected Gaia Catalogue, including the effect of observational errors. We statistically analyse this simulated Gaia data to better understand what can be obtained from the Gaia astrometric mission. This catalogue is used to investigate the potential yield in astrometric, photometric, and spectroscopic information and the extent and effect of observational errors on the true Gaia Catalogue. This article is a follow-up to Robin et al. (A\&A 543, A100, 2012), where the expected Gaia Catalogue content was reviewed but without the simulation of observational errors.}
{We analysed the Gaia Object Generator (GOG) catalogue using the Gaia Analysis Tool (GAT), thereby producing a number of statistics about the catalogue.}
{A simulated catalogue of one billion objects is presented, with detailed information on the 523 million individual single stars it contains. Detailed information is provided for the expected errors in parallax, position, proper motion, radial velocity, and the photometry in the four Gaia bands and for the physical parameter determination including temperature, metallicity, and line of sight extinction. }
{}

\maketitle

\section{Introduction} 

Gaia, a cornerstone ESA mission, launched in December 2013, will produce the fullest 3D galactic census to date, and it is expected to yield a huge advancement in our understanding of the composition, structure, and evolution of the Galaxy \citep{Gaia}.  
Through Gaia's photometric instruments, object detection up to $G=20$ mag will be possible (see \cite{Gmag} for a definition of $G$ magnitude), including measurements of positions, proper motions, and parallaxes up to micro arcsecond accuracy. The on-board radial velocity spectrometer will provide radial velocity measurements for stars down to a limit of $G_{RVS}=17$ mag.  With low-resolution spectra providing information on effective temperature, line of sight extinction, surface gravity, and chemical composition, Gaia will yield a detailed catalogue that contains roughly 1\% of the entire galactic stellar population. 

Gaia will represent a huge advance on its predecessor, Hipparcos \citep{Hipparcos}, both in terms of massive increases in precision and in the numbers of objects observed. Thanks to accurate observations of large numbers of stars of all kinds, including rare objects, large numbers of Cepheids and other variable stars, and direct parallax measurements for stars in all galactic populations (thin disk, thick disk, halo, and bulge), Gaia data is expected to have a strong impact on luminosity calibration and improvement of the distance scale. This, along with applications to studies of galactic dynamics and evolution and of fields ranging from exoplanets to general relativity and cosmology, Gaia's impact is expected to be significant and far reaching.

During its five years of data collection, Gaia is expected to transmit some 150 terabytes of raw data to
Earth, leading to production of a catalogue of $10^9$ individual objects. After on-ground processing, the full database is expected to be in the range of one to two petabytes of data. 
Preparation for acquiring this huge amount of data is essential. Work has begun to model the expected output of Gaia in order to predict the content of the Gaia Catalogue, to facilitate the production of tools required to effectively validate the real data before publication, and to analyse the real data set at the end of the mission.

To this end, the Gaia Data Processing and Analysis Consortium (DPAC) has been preparing a set of simulators, including a simulator called the Gaia Object Generator (GOG), which simulates the end-of-mission catalogue, including observational errors. Here a full description of GOG is provided, including the models assumed for the performance of the Gaia satellite and an overview of its simulated end-of-mission catalogue. A selection of statistics from this catalogue is provided to give an idea of the performance and output of Gaia.

In Sect. \ref{sec:gaia}, a brief description of the Gaia instrument and an overview of the current simulation effort is given, followed by definitions of the error models assumed for the performance of the Gaia satellite in Sec. \ref{sec:simulations}. In Sec. \ref{sec:methods}, the methods used for searching the simulated catalogue and generating statistics are described. In Sec. \ref{sec:results}, we present the results of the full sky simulation, broken up into sections for each parameter in the catalogue and specific object types of interest. Finally, in Sec. \ref{sec:conclusions} we provide a summery and conclusions.

\section{The Gaia satellite}
\label{sec:gaia}

The Gaia space astrometry mission will map the entire sky in the visible $G$ band over the course of its five-year mission. Located at Lagrangian point L2, Gaia will be constantly and smoothly spinning. It has two telescopes separated by a basic angle of $106.5^\circ$. Light from stars that are observed in either telescope is collected and reflected to transit across the Gaia focal plane. 

The Gaia focal plane can be split into several main components. The majority of the area is taken up by CCDs for the broad band $G$ magnitude measurements in white light, used in taking the astrometry measurements. Second, there is a pair of low-resolution spectral photometers, one red and one blue, producing low-resolution spectra with integrated magnitude $G_{RP}$ and $G_{BP}$, respectively. Finally, there is a radial velocity spectrometer observing at near-infrared, with integrated magnitude $G_{RVS}$. The magnitudes $G$, $G_{RP}$, and $G_{BP}$ will be measured for all Gaia sources ($G\le 20$), whereas $G_{RVS}$ will be measured for objects up to $G_{RVS}\le 17$ magnitude. For an exact definition of the Gaia focal plane and the four Gaia bands, see Figs. 1 and 3 of \cite{Gmag}.

The motion of Gaia is complex, with rotations on its own spin axis occurring every six hours. This spin axis is itself precessing, and is held at a constant $45^\circ$ degrees from the Sun. From its position at L2, Gaia will orbit the Sun over the course of a year. Thanks to the combination of these rotations, the entire sky will be observed repeatedly. The Gaia scanning law gives the number of times a region will be re-observed by Gaia over its five-year mission, and comes from this spinning motion of the satellite and its orbit around the Sun. Objects in regions with more observations have greater precision, while regions with fewer repeat observations have lower precision. The average number of observations per object is 70, although it can be as low as a few tens or as high as 200. 

\section{Simulations}
\label{sec:simulations}
Simulation of many aspects of the Gaia mission has been carried out in order to test and improve instrument design, data reduction algorithms, and tools for using the final Gaia Catalogue data. The Gaia Simulator is a collection of three data generators designed for this task: the Gaia Instrument and Basic Image Simulator  \citep[GIBIS,][]{gibis}, the Gaia System Simulator  \citep[GASS,][]{masana}, and the Gaia Object Generator (GOG, described here). Through these three packages, the production of the simulated Gaia telemetry stream, observation images down to pixel level and intermediate or final catalogue data is possible. 
 
\subsection{Gaia Universe Model Snapshot and the Besan\c{c}on galaxy model}

One basic component of the Gaia Simulator is its Universe Model (UM), which is used to create object catalogues down to a particular limiting magnitude (in our case $G=20$ mag for Gaia). For stellar sources, the UM is based on the Besan\c{c}on galaxy model \citep{Besancon}. This model simulates the stellar content of the Galaxy, including stellar distribution and a number of object properties. It produces stellar objects based on the four main stellar populations (thin disk, thick disk, halo, and bulge), each population with its own star formation history and stellar evolutionary models. Additionally, a number of object-specific properties are also assigned to each object, dependent on its type. Possible objects are stars (single and multiple), nebulae, stellar clusters, diffuse light, planets, satellites, asteroids, comets, resolved galaxies, unresolved extended galaxies, quasars, AGN, and supernovae. Therefore, the UM is capable of simulating almost every object type that Gaia can potentially observe. It can therefore construct simulated object catalogues down to Gaia's limiting magnitude. 

Building on this, the UM creates for any time, over any section of the sky (or the whole sky), a set of objects with positions and assigns each a set of observational properties \citep{robin2011}. These properties include distances, apparent magnitudes, spectral characteristics, and kinematics. 

Clearly the models and probability distributions used in order to create the objects with their positions and properties are highly important in producing a realistic catalogue. The UM has been designed so that the objects it creates fit as closely as possible to observed statistics and to the latest theoretical formation and evolution models \citep{Besancon}. For a statistical analysis of the underlying potentially observable population (with $G\le 20$ mag) using the Gaia UM without satellite instrument specifications and error models, see \cite{robin2011}.  

\subsection{The Gaia Object Generator}
The GOG is capable of transforming this UM catalogue into Gaia's simulated intermediate and final catalogue data. This is achieved through the use of analytical and numerical error models to create realistic observational errors in astrometric, photometric, and spectroscopic parameters \citep{GOG}. In this way, GOG transforms `true' object properties from the UM into `observed' quantities that have an associated error that depends on the object's properties, Gaia's instrument capabilities, and the type and number of observations made.

\subsection{Error models}
DPAC is divided into a number of coordination units (CUs), each of which specialises in a specific area. In GOG we have taken the recommendations from the various CUs in order to include the most complete picture of Gaia performance as possible.  The CUs are divided into the following areas: CU1, system architecture; CU2, simulations; CU3, core processing (astrometry); CU4, object processing (multiple stars, exoplanets, solar system objects, extended objects); CU5, photometric processing; CU6, spectroscopic processing; CU7, variability processing; CU8, astrophysical parameters; and CU9, catalogue access. 

Models for specific parameters have been provided by the various CUs, and only an outline is given here. In the following description, \emph{true} refers to UM data (without errors),  \emph{epoch}  to simulated individual observations (including errors), and \emph{observed}  to the simulated observed data for the end of the Gaia mission (including standard errors). Throughout, \emph{error} refers to the formal standard error on a measurement.

\subsubsection{Astrometric error model}
\label{sec:endofmissiondatamodels}
The formal error on the parallax, $\sigma_{\varpi}$, is calculated following the expression:
        \begin{equation} \sigma_{\varpi} = m \cdot g_{\varpi} \cdot \sqrt{\frac{\sigma_{\eta}^2}{N_{\text{eff}}}+ \frac{\sigma_{\text{cal}}^2}{N_{\text{transit}}}} \end{equation}

\begin{itemize}
    \item $\sigma_\eta$ is the CCD centroid positioning error. It uses the Cramer Rao (CR) lower bound in its discrete form, which defines the best possible precision of the maximum likelihood centroid location estimator. The CR lower bound requires the line spread function (LSF) derivative for each sample\footnote{In Gaia, a sample is defined as a set of individual pixels.}, the background, the readout noise, and the source integrated signal. 
        \item $m$ is the contingency margin that is used to take scientific and environmental effects into account, for example: uncertainties in the on-ground processing, such as uncertainties in relativistic corrections and solar system ephemeris; effects such as having an imperfect calibrating LSF; errors in estimating the sky background; and other effects when dealing with real stars. The default value assumed for the Gaia mission has been set by ESA as 1.2 and is used in GOG.
        \item $g_{\varpi} = 1.47 / \text{sin} \xi$  is a geometrical factor where $\xi$ is known as the solar aspect angle, with a value of 45$\degree$.
        \item $N_{\text{eff}}$ is the number of elementary CCD transits $(N_{\text{strip}} \times N_{\text{transit}})$. 
        \item $N_{\text{transit}}$ is the number of field of view transits. 
        \item $N_{\text{strip}}$ is the number of CCDs in a row on the Gaia focal plane. It has a value of 9, except for the row that includes the wavefront sensor, which has 8 CCDs. 
         \item $\sigma_{\text{cal}}$ is the calibration noise. A constant value of 5.7 $\mu$as has been applied. This takes into account that the end-of-mission precision on the astrometric parameters not only depends on the error due to the location estimation with each CCD. There are calibration errors from the CCD calibrations, the uncertainty of the attitude of the satellite and the uncertainty on the basic angle. 
    
\end{itemize}

We are enabling the activation of gates, as described in the Gaia Parameter Database. The Gaia satellite will be smoothly rotating and will constantly image the sky by collecting the photons from each source as they pass along the focal plane. The total time for a source to pass along the focal plane will be 107 seconds, and the electrons accumulated in the a CCD pixel will be passed along the CCD at the same rate as the source. To avoid saturation for brighter sources, and the resulting loss of astrometric precision, gates can be activated that limit the exposure time. Here we are using the default gating system, which could change during the mission.

Following \cite{gaiascience} (see also http://www.cosmos.esa.int/web/gaia/science-performance), we have assumed that the errors on the positional coordinates at the mean epoch (middle of the mission), and the error in the proper motion coordinates can be given respectively by

\begin{itemize}
        \item $\sigma_{\alpha} = 0.787 \sigma_{\varpi} $
        \item $\sigma_{\delta} = 0.699 \sigma_{\varpi} $
        \item $\sigma_{\mu_\alpha} = 0.556 \sigma_{\varpi} $
        \item $\sigma_{\mu_\delta} = 0.496 \sigma_{\varpi} $
\end{itemize}

\subsubsection{Photometric error model}

GOG uses the single CCD transit photometry error $\sigma_{\text{p,j}}$ \citep{Gmag} defined as

        \begin{equation}\sigma_{\text{p,j}} [\text{mag}] = 2.5 \cdot \log_{10} (e) \cdot \frac{\sqrt{f_{\text{aperture}} \cdot s_\text{j} + (b_\text{j} +r^2) \cdot n_s \cdot (1+ \frac{n_\text{s}}{n_\text{b}})}}{f_{\text{aperture}} \cdot s_\text{j}} \end{equation} to compute either the epoch errors or the end-of-mission errors.

We assume, following an `aperture photometry' approach, that the object flux $s_\text{j}$ is measured in a rectangular `aperture' (window) of $n_\text{s}$ along-scan object samples. The sky background $b_\text{j}$ is assumed to be measured in $n_\text{b}$ background samples, and $r$ is CCD readout noise.
The $f_{\text{aperture}}\cdot s_\text{j}$ is expressed in units of photo-electrons ($e^-$), and denotes the object flux in photometric band $j$ contained in the `aperture' (window) of $n_\text{s}$ samples, after a CCD crossing. The number $f_{\text{aperture}}$  thus represents the fraction of the object flux measured in the aperture window. 

For the epoch and end-of-mission data, the same expression is used for the standard deviation calculation
\begin{equation}  \sigma_{\text{G,j}} = m \cdot \sqrt { \frac{\sigma^2_{\text{p,j}} + \sigma_{\text{cal}}^2} {N_{\text{eff}}} } 
\end{equation}  
where $N_{\text{eff}}$ is the number of elementary CCD transits $(N_{\text{strip}} \times N_{\text{transit}})$, with $(N_{\text{strip}}=1$ and $N_{\text{transit}})=1$ for epoch photometry and equal to the number of transits for the end-of-mission photometry. The calibration noise $\sigma_{\text{cal}}$ has a fixed value of $\sigma_{\text{cal}} = 30$ mmag. Use of a fixed value of the centroiding error is possible because this error is only relevant for brighter stars, because their centroiding errors are smaller than the calibration error.

\subsubsection{Radial velocity spectrometer errors}

CU6 tables (see Table \ref{tab:RVS}) using the stars' physical parameters and apparent magnitude are used to obtain $\sigma_{V_\text{r}}$. They were computed following the prescriptions of \cite{KAT04} and later updates. Those tables have been provided for one and 40 field-of-view transits, therefore the value for 40 transits is used here to calculate the average end-of-mission errors in RVS. 

Given the information on the apparent Johnson V magnitude and the atmospheric parameters of each star (from the UM), we select from Table \ref{tab:RVS} the closest spectral type and return the corresponding radial velocity error. Since Table \ref{tab:RVS} is given for [Fe/H]=0 alone, a correction is made on the apparent magnitude in order to take different metallicities into account: for each variation in metallicity of $\Delta$ [Fe/H]=−1.5 dex, the magnitude is increased by $V$ = 0.5 mag.

We have set a lower limit on the wavelength calibration error, giving a lower limit on the radial velocity error of 1 km$\cdot$s$^{-1}$. For the faintest stars the spectra will be of poor quality and will not contain enough information to enable accurate estimation of the radial velocity. Owing to the limited bandwidth in the downlink of Gaia data to Earth, poor quality spectra will not be transmitted. We therefore set an upper limit on the radial velocity error of 20 km$\cdot$s$^{-1}$, beyond which we assume that there will be no data. The exact point at which the data will be assumed to have too low a quality is still unknown.

\begin{table*}\centering
    \begin{tabular}{@{}lllllllllllllllllll@{}}\toprule 
\backslashbox{Type}{$V$}  &  8.5  &  9.0  &  9.5  &  10  &  10.5  &  11.5  &  12  &  12.5  &  13  &  13.5  &  14  &  14.5  &  15  &  15.5  &  16  &  16.5  &  17  &  17.5 \\ \midrule
B0V   &  1.2 &  1.6 &  2   &  2.7 &  3.8 &  6.8 &  9.7 &  14.5 &  24.8  &  n   &  n    &  n  &  n  &  n  &  n  &  n  &  n  &  n  \\
B5V   &  1   &  1.1 &  1.4 &  1.9 &  2.5 &  5.1 &  6.9 &  10   &  15.3  & 24.1 &  n    &  n  &  n  &  n  &  n  &  n  &  n  &  n  \\
A0V   &  1   &  1   &  1   &  1   &  1   &  1.3 &  1.8 &  2.6  &  3.9 &  5.7   &  8.6  &  14.6  &  32.5  &  n  &  n  &  n  &  n &  n \\
A5V   &  1   &  1   &  1   &  1   &  1   &  1   &  1   &  1.3  &  2   &  4.2   &  6.9  &  11.1  &  20.1  &  n  &  n  &  n  &  n &  n \\
F0V   &  1   &  1   &  1   &  1   &  1   &  1   &  1   &  1    &  1.5 &  2.1   &  3.2  &  5.3   &  7.8  &  12.7 &  23.4 &  n  &  n  &  n \\
G0V   &  1   &  1   &  1   &  1   &  1   &  1   &  1   &  1    &  1   &  1.4   &  2.1  &  3     &  4.8  &  7.9  &  12.4 &  19.6 &  n  &  n \\
G5V   &  1   &  1   &  1   &  1   &  1   &  1   &  1   &  1    &  1   &  1.2   &  1.9  &  2.8   &  4.4  &  6.3  &  10.1 &  17.6 &  n  &  n \\
K0V   &  1   &  1   &  1   &  1   &  1   &  1   &  1   &  1    &  1   &  1.1   &  1.4  &  2.1   &  3.3  &  5.1  &  8.1  &  12.6 &  24.9  &  n \\
K4V   &  1   &  1   &  1   &  1   &  1   &  1   &  1   &  1    &  1   &  1     &  1.1  &  1.6   &  2.7  &  3.6  &  5.2  &  8.4  &  14.5  &  30 \\
K1III &  1   &  1   &  1   &  1   &  1   &  1   &  1   &  1    &  1   &  1     &  1    &  1.2   &  1.8  &  2.7  &  4.2  &  6.8  &  10.3  &  18 \\

 \bottomrule
    \end{tabular}
    \caption{The average end-of-mission formal error in radial velocity with an assumed average of 40 field-of-view transits, in km$\cdot$s$^{-1}$, for each spectral type. The numbers in the top row are Johnson apparent V magnitudes. Fields marked by ``n'' are assumed to be too faint to produce spectra with sufficient quality for radial velocity determination. Stars with these magnitudes will have no radial velocity information.}
    \label{tab:RVS}

\end{table*}

\subsubsection{Physical parameters}

GOG uses the stellar parametrisation performance given by CU8 to calculate error estimations for effective temperature, line-of-sight extinction, metallicity, and surface gravity. The colour-independent extinction parameter $A_0$ is used in preference to the band specific extinctions $A_V$ or $A_G$, because $A_0$ is a property of the interstellar medium alone \citep{cbj11}. CU8 use three different algorithms to calculate physical parameters using spectrophotometry (see \cite{CU8}). 

It should be noted that the errors calculated here are calculated only as a function of apparent magnitude. However, as described in \cite{CU8}, there are clear dependencies on the spectral type of the star, because some star types may or may not exhibit spectral features required for parameter determination. Additionally, \cite{CU8} report a strong correlation between the estimation of effective temperature and extinction. This correlation is not simulated in GOG.
Following the recommendation of CU8, calculating errors of physical parameters depends on apparent magnitude and is split into two cases, objects with $A_0<1$ mag and $A_0\geq 1$ mag. 

In GOG, $\sigma_{T_{\text{eff}}}$, $\sigma_{A_0}$, $\sigma_{Fe/H}$ and $\sigma_{\log g}$ are calculated from a Gamma distribution, with shape parameter $\alpha$ and scale parameter $\theta$ :
\begin{equation} f(\sigma; \alpha, \theta) = \frac{1}{\Gamma(\sigma)\theta^\alpha} \sigma^{\alpha -1} e^{-\frac{x}{\theta} }   
\end{equation}
where $\alpha$ and $\theta$ are obtained from the following expressions, which have been calculated to give each $\sigma$ a close approximation to the CU8 algorithm results. A gamma distribution was selected for ease of implementation and for its ability to statistically recreate the CU8 results to a reasonable approximation. A gamma distribution is also only non-zero for positive values of sigma. This is essential when modelling errors because, of course, it is impossible to have a negative error.

\begin{itemize}
\item For stars with $A_0<1$ mag:
\[\alpha_{A_0} = 0.204 - 0.032  G + 0.001  G^2 \]
\[\alpha_{\log g} = 0.151 - 0.019   G + 0.001   G^2 \]
\[\alpha_{Fe/H} = 0.295  - 0.047  G + 0.002   G^2 \]
\[\alpha_{T_{\text{eff}}} = 78.2  - 10.3   G + 0.46  G^2 \]
\[\theta_{A_0} = 0.084  \]
\[\theta_{\log g} = 0.160\]
\[\theta_{Fe/H} = 0.121 \]
\[\theta_{T_{\text{eff}}} = 28.2. \]

\item For stars with $A_0 \geq 1$ mag:
\[\alpha_{A_0} = 0.178 - 0.026   G + 0.001   G^2 \]
\[\alpha_{\log g} = 0.319  - 0.044   G + 0.002   G^2 \]
\[\alpha_{Fe/H} = 0.717  - 0.115  G + 0.005   G^2 \]
\[\alpha_{T_{\text{eff}}} = 67.3 - 7.85  G + 0.35  G^2 \]
\[\theta_{A_0} = 0.096\]
\[\theta_{\log g} = 0.179\]
\[\theta_{Fe/H} = 0.353\]
\[\theta_{T_{\text{eff}}} = 33.5.\]

\end{itemize}

The Gamma distributions thus obtained for each parameter are used to generate a formal error for each parameter for each individual star, aiming to statistically (but not individually) reproduce the results that will be obtained by the application of the CU8 algorithms and then included in the Gaia Catalogue.

It should be noted that in the stellar parametrisation algorithms used in \cite{CU8}, a degeneracy is reported between extinction and effective temperature owing to the lack of resolved spectral lines only sensitive to effective temperature. In GOG, this degeneracy has not been taken into account, and the precisions of each of the four stellar parameters is simulated independently.

Additionally, the results of \cite{CU8} have recently been updated, and \cite{cbj13} gives the latest results regarding the capabilities of physical parameter determination. This latest paper has not been included in the current version of GOG.

\subsection{Limitations}

In the present paper, only the results for single stars are given in detail. All of the figures and tables represent the numbers and statistics of only individual single stars, excluding all binary and multiple systems. Since the performance of the Gaia satellite is largely unknown for binary and multiple systems, the implementation into GOG of realistic error models has not yet been possible. While the results presented in Sec. \ref{sec:results} are expected to be reliable under current assumptions for the performance of Gaia, the real Gaia Catalogue will differ from these results thanks to the presence of binary and multiple systems. By removing binaries from the
latter, direct comparison of the results presented here with the real Gaia Catalogue  will not be possible because of the presence of unresolved binaries, which are difficult to detect. As a simulator, GOG relies heavily on all inputs and assumptions supplied both from the UM or from the Gaia predicted performance and error models. 

In our simulations we used an exact cut at $G=20$ mag, beyond which no stars are observed. In reality, in regions of low density observations of stars up to 20.5 mag could be possible. Inversely, very crowded regions may not be complete up to 20 mag, or the numbers of observations per star over the five-year mission may be reduced in these regions.

There is no simulation of the impact of crowding on object detection or the detection of components in binary and multiple systems. This can lead to unrealistic quality in all observed data in the most crowded regions of the plane of the Galaxy, to overestimates for star counts in the bulge, and to a lack of features related to the disk and bulge in Figs. \ref{fig:parerrormap} and \ref{fig:photometryMaps}.

Additionally, GOG uses the nominal Gaia scanning law to calculate the number of field of view transits per object over the five years of the mission while Gaia is operating in normal mode. There will be an additional one-month period at the start of the mission using an ecliptic pole scanning law, and this has not been taken into account. It may lead to a slight underestimation of the number of transits, and therefore a slight overestimation of errors, for some stars near the ecliptic poles.

There is the possibility that the Gaia mission will be extended above the nominal five-year mission. Since this idea is under discussion and has not yet been confirmed or discarded, we only present results for the Gaia mission as originally planned. 

If the length of the mission is increased, the number of field-of-view transits will increase, and the precision per object will improve. If the proposal is accepted, the GOG simulator could be used to provide updated statistics for the expected catalogue without extensive modification.

\section{Methods and statistics}
\label{sec:methods}

Considering current computing capabilities, it is not straightforward to make statistics and visualisations when dealing with catalogues of such a large size. A specific tool has been created which is capable of extracting information and visualising results, with excellent scalability allowing its use for huge datasets and distributed computing systems. 

The Gaia Analysis Tool (GAT) is a data analysis package that allows, through three distinct frameworks, generation of statistics, validation of data, and generation of catalogues. It currently handles both UM- and GOG-generated data, and could be adapted to handle other data types.

Every statistical analysis is performed by a Statistical Analysis Module (SAM), with several grouped into a single XML file as an input to GAT. Each SAM can contain a set of filters, enabling analysis of specific subsets of the data. This allows the production of a wide range of statistics for objects satisfying any number of specific user-defined criteria or for the catalogue as a whole. 

GAT creates a number of different statistics outputs including histograms, sky density maps and HR diagrams. After the GAT execution, statistics output are stored to either generate a report or to be analysed using the GAT Displaying tool.

Because we have information from not only observations of a population but also of the observed population itself, comparison is possible between the simulated Gaia Catalogue and the simulated `true' population, allowing large scope for investigating the precision\footnote{Here we assume the standard definition of accuracy and precision: accuracy is the closeness of a result (or set) to the actual value, i.e. it is a measure of systematics or bias. Precision is the extent of the random variability of the measurement, i.e. what is called observational errors above.} of the observations and differences between the two catalogues. Clearly this is only possible with simulated data and cannot be attempted with the true Gaia Catalogue, so it is an effective way to investigate the possible extent and effect of observational errors and selection bias on the real Gaia Catalogue, where this kind of comparison is not possible.

GOG can be used in the preparations for validating the true Gaia Catalogue, by testing the GOG catalogue for accuracy\footnotemark[2] and precision. In special cases, observational biases could even be implemented into the code to allow thorough testing of validation methods.

\begin{figure*}[t]
\begin{center}
\includegraphics[scale=.5]{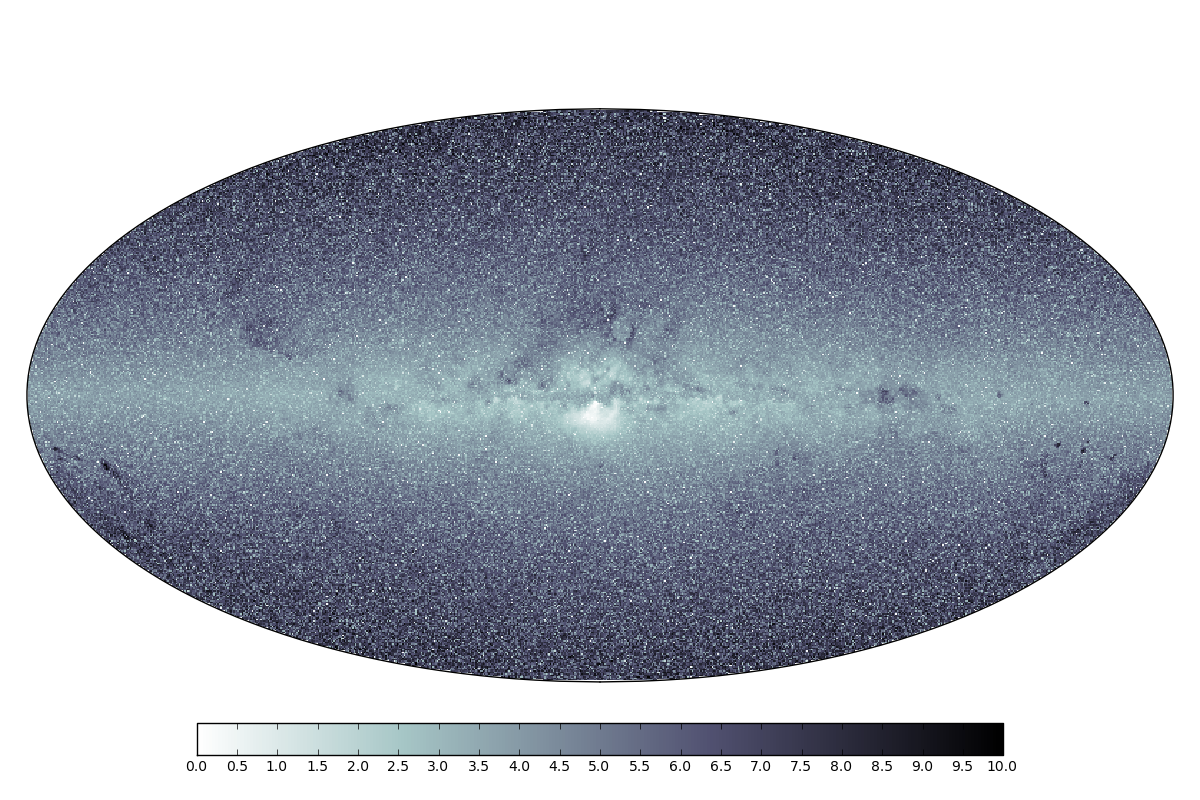} \caption{Skymap of total integrated flux over the Milky Way, in the $G$ band. The colour bar represents a relative scale, from maximum flux in white to minimum flux in black. The figure is plotted in galactic coordinates with the galactic-longitude orientation swapped left to right.}
\label{fig:TotalFlux}
\end{center}    
\end{figure*}

\section{Results}
\label{sec:results}
The GOG simulator has been used to generate the simulated final mission catalogue for Gaia, down to magnitude $G=20$. The simulation was performed on the MareNostrum super computer at the Barcelona Supercomputing Centre (Centre Nacional de Supercomputaci\'{o}), and it took 400 thousand CPU hours. An extensive set of validations and statistics have been produced using GAT to validate performance of the simulator. Below we include a subset of these statistics for the most interesting cases to give an overview of the expected Gaia Catalogue.

\subsection{General}

In total, GOG has produced a catalogue of about one billion objects, consisting of 523 million individual single stars and 484 million binary or multiple systems. The total number of stars, including the components of binary and multiple systems is 1.6 billion. The skymap of the total flux detected over the entire five-year mission is given in Fig. \ref{fig:TotalFlux}. Although GOG can produce extragalactic sources, none have been simulated here.

The following discussion is split into sections for different types of objects of interest. Section \ref{sec:Stars} covers all galactic stellar sources. Section \ref{sec:Variables} covers all variable objects. Section \ref{sec:Params} is a discussion of physical parameters estimated by Gaia. All objects in these sections are within the Milky Way.

To make the presentation of performance as clear as possible, all binary and multiple systems and their components have been removed from the following statistics. This is due to complicating effects that arise when dealing with binary systems, some of which GOG is not yet capable of correctly simulating; for example, GOG does not yet contain an orbital solution in its astrometric error models, and the effects of unresolved systems on photometry and astrophysical parameter determination have not yet been well determined. Therefore, the numbers presented are only for individual single stars and do not include the full one billion objects simulated. Of the single stars presented in this paper, 74 million are within the radial velocity spectrometer  magnitude range.

Table \ref{tab:summary table} gives the mean and median error for each of the observed parameters discussed in this paper, along with the upper and lower 25\% quartile.

\begin{figure*}[ht]
\centering
\subfigure{%
\includegraphics[width=0.45\linewidth]{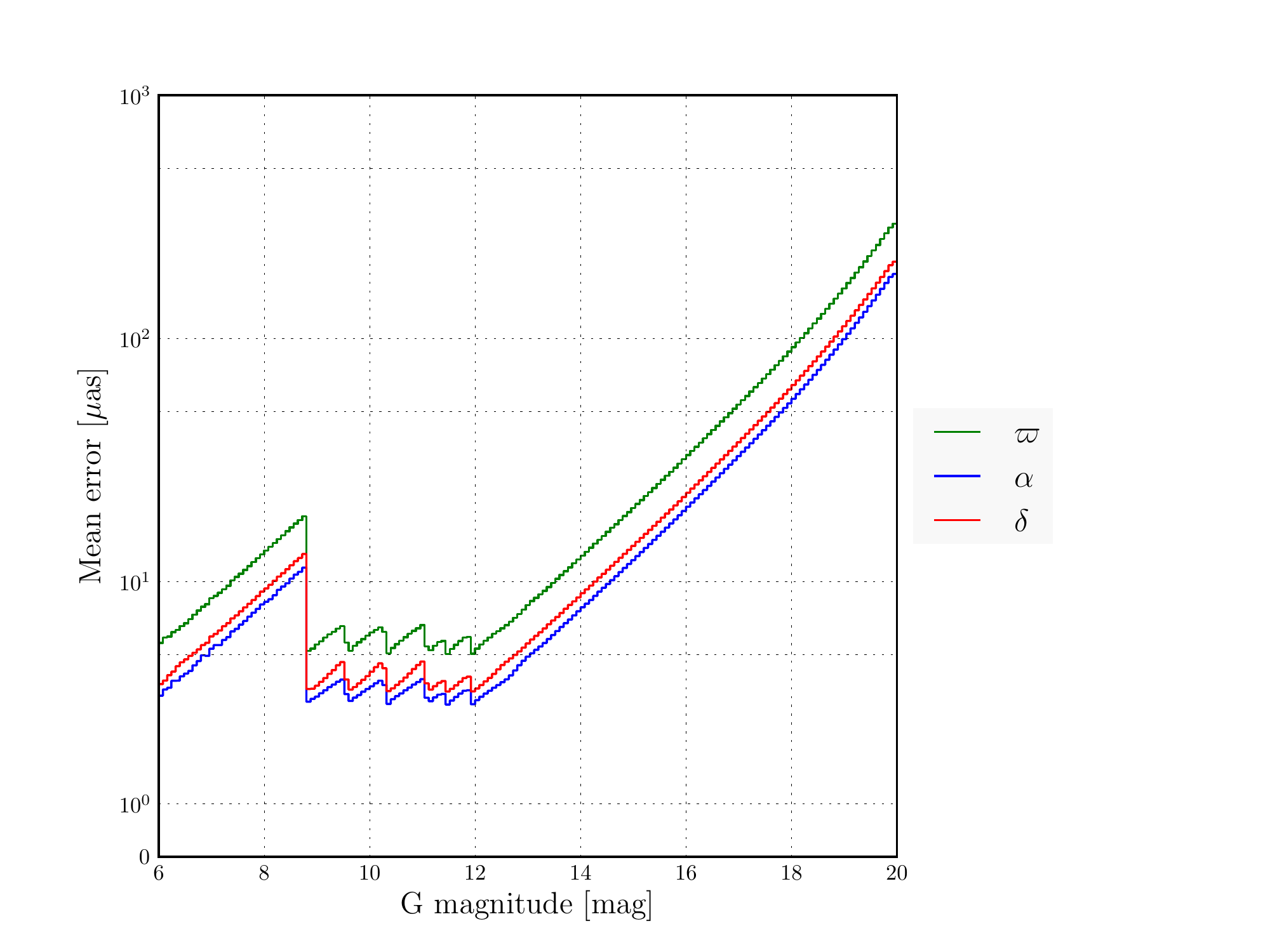}}
\quad
\subfigure{%
\includegraphics[width=0.45\linewidth]{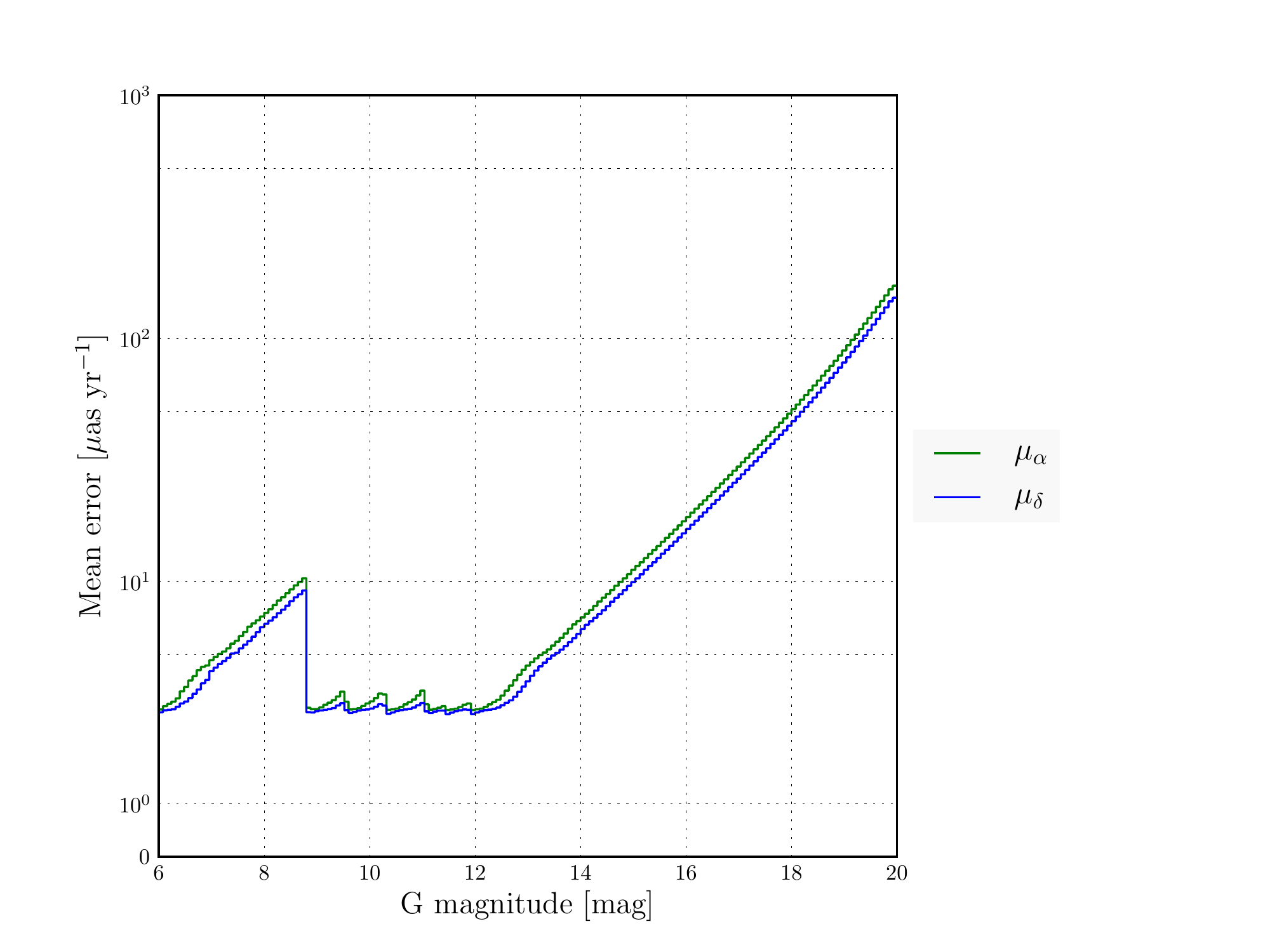}}
\subfigure{%
\includegraphics[width=0.45\linewidth]{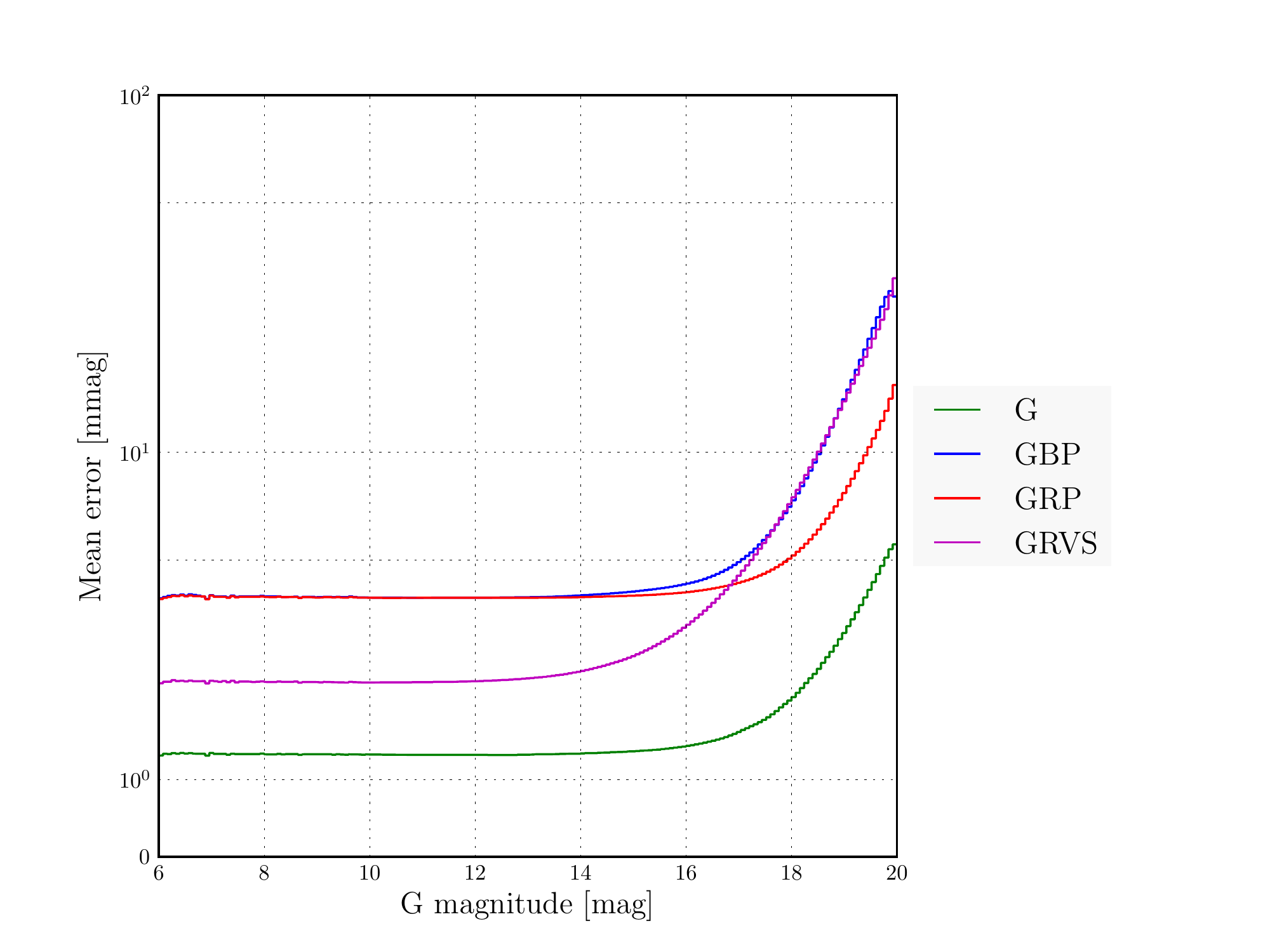}}
\quad
\subfigure{%
\includegraphics[width=0.45\linewidth]{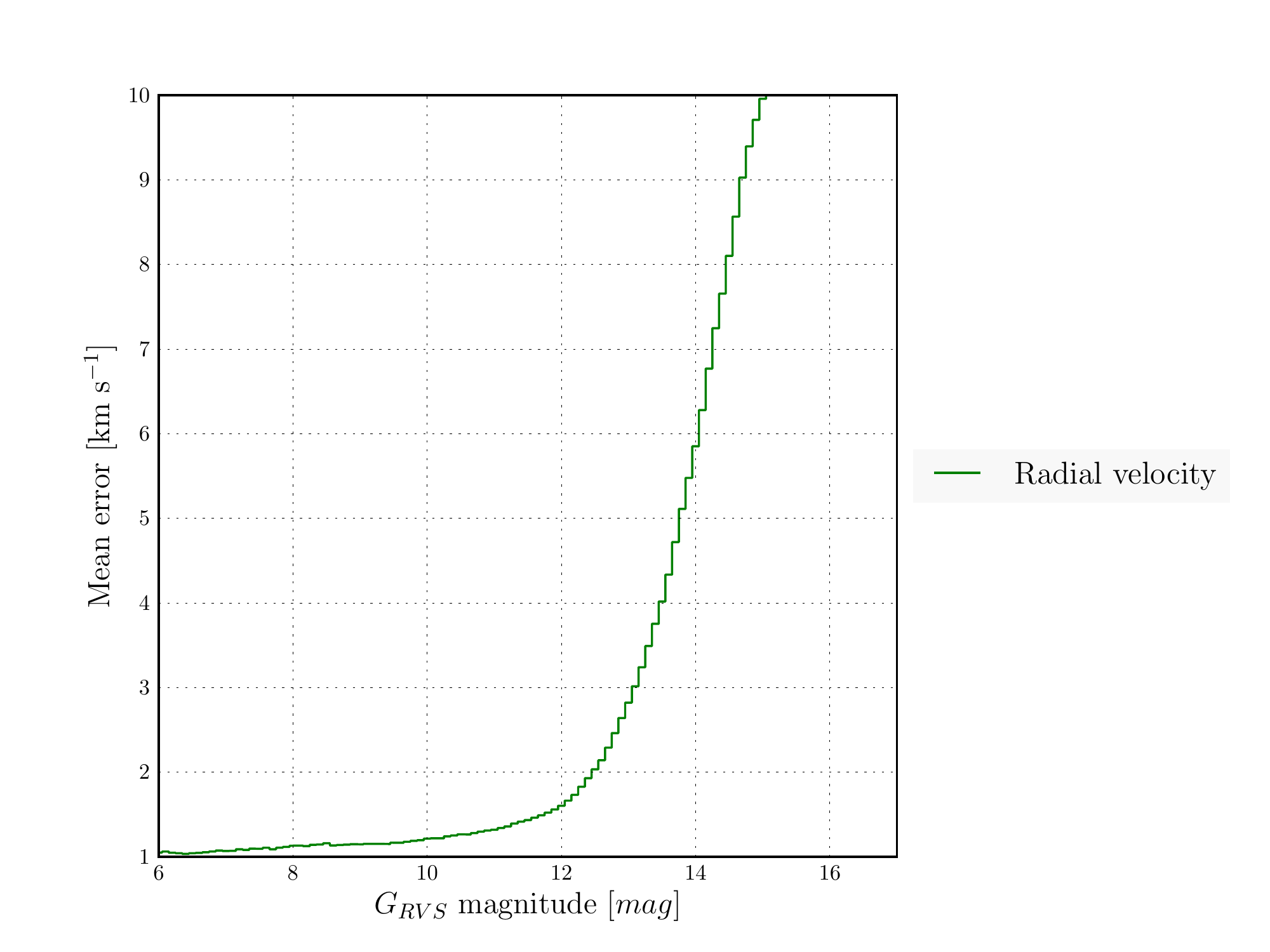}}

    \caption{Mean end-of-mission error as a function of $G$ magnitude for parallax, position, proper motion, and photometry in the four Gaia passbands. Additionally the mean end-of-mission error in radial velocity as a function of $G_{RVS}$ magnitude.}
    \label{fig:MeanErrors}
\end{figure*}

In Fig. \ref{fig:MeanErrors}, the mean error for parallax, position, proper motion, and photometry in the four Gaia bands are given as a function of $G$ magnitude. Also, the mean error in radial velocity is given as a function of G$_{RVS}$ magnitude. The sharp jumps in the mean error in astrometric parameters between 8 and 12 mag are due to the activation of gates for the brighter sources in an attempt to prevent CCD saturation (see Sec. \ref{sec:endofmissiondatamodels}).

\begin{table}\centering
    \begin{tabular}{@{}lllllll@{}}\toprule 
        
     Standard error &  LQ &    Median & Mean &  UQ   \\ \midrule
Parallax ($\mu$as)                &  80     & 140     &   147  &  210    \\
$\alpha^*$ ($\mu$as)              &  40     &  80     &   91   &  130    \\
$\delta$ ($\mu$as)                &  50     &  100    &   103  &  150    \\
$\mu_\alpha$ ($\mu$as$\cdot$yr$^{-1}$)  &  40     &  80     &   82   &  120    \\        
$\mu_\delta$ ($\mu$as$\cdot$yr$^{-1}$)  &  40     &  70     &   73   &  110    \\
$G$ (mmag)                        &  2      &   3     &   3.0  &    4    \\
$G_{BP}$ (mmag)                   &  6      &   11    &   14.6 &   19    \\
$G_{RP}$ (mmag)                   &  5      &  7      &   7.7  &   10    \\
$G_{RVS}$ (mmag)                  &  6      &  11     &   13.2 &   18    \\
Radial velocity (km$\cdot$s$^{-1}$)      &  3      &  7      &   8.0  &   13    \\
 Extinction (mag)                 &  0.16   &  0.21   &   0.21 &  0.26   \\
Metallicity (Fe/H)                &  0.46   &  0.57   &   0.57 &  0.73   \\
Surface gravity ($\log g$)        &  0.34   &  0.35   &   0.45 &  0.58   \\
Effective temperature (K)         &  280    &  350    &   388  &  530    \\        
        \bottomrule
    \end{tabular}
    \caption{Mean and median value of the end-of-mission error in each observable, along with the upper (UQ) and lower (LQ) 25\% quartile. Since the error distributions are not symmetrical, the mean value should not be used directly, and is given only to give an idea of the approximate level of Gaia's precision. The median $G$ magnitude of all single stars is 18.9 mag.}
    \label{tab:summary table}

\end{table}

\subsection{Stars}\label{sec:Stars}
\subsubsection{Parallax}

The distribution of parallax measurements for all single stars is given in Fig. \ref{fig:parallax}. The mean parallax error for all single stars is 147.0 $\mu$as. The number of single stars falling below three relative parallax error limits is given in Table \ref{tab:relerrorpiSpectraltype} for each spectral type and in Table \ref{tab:relerrorpiLumClass} for each luminosity class. For those interested in a specific type of star, Table \ref{tab:relerrorpiByAll} gives the full breakdown of the number of single stars falling below three relative parallax error limits for every spectral type and luminosity class. The distribution of parallax errors is given for each stellar population in Fig. \ref{fig:errorpibypop} and for each spectral type in Fig. \ref{fig:errorpispectraltype}. The relative parallax error $\sigma_\varpi/\varpi$ is given in Fig. \ref{fig:relerrorpiSpectralType} for stars split by spectral type.

\begin{figure}
\begin{center}
\includegraphics[width=\linewidth]{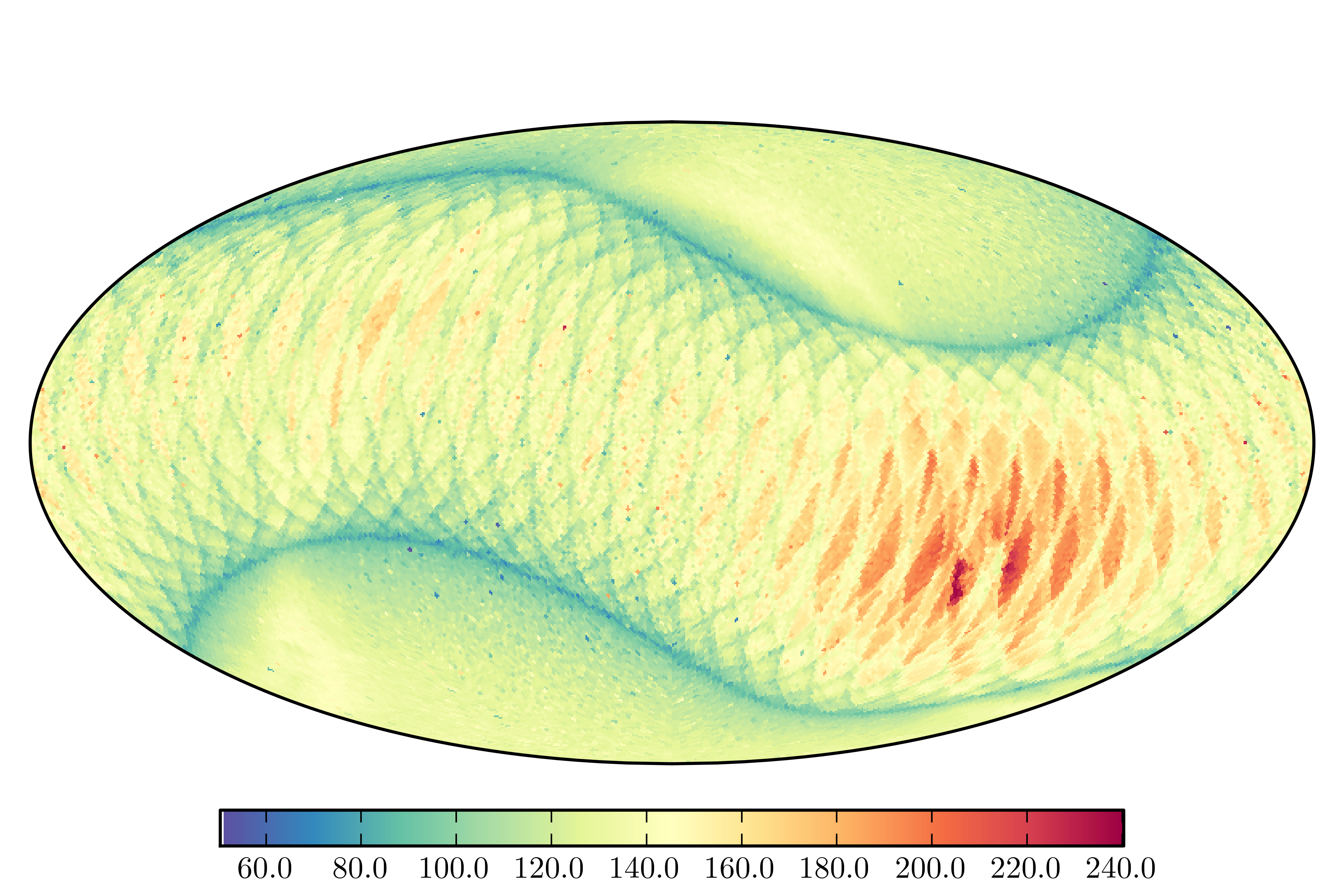} 
\caption{Sky map (healpix) of mean parallax error for all single stars in equatorial coordinates. Colour scale is mean parallax error in $\mu$as. The red area is the location of the bulge. }
\label{fig:parerrormap}
\end{center}    
\end{figure}

\begin{figure*}[ht]
\centering
\begin{minipage}[b]{0.45\linewidth}
\includegraphics[width=\linewidth]{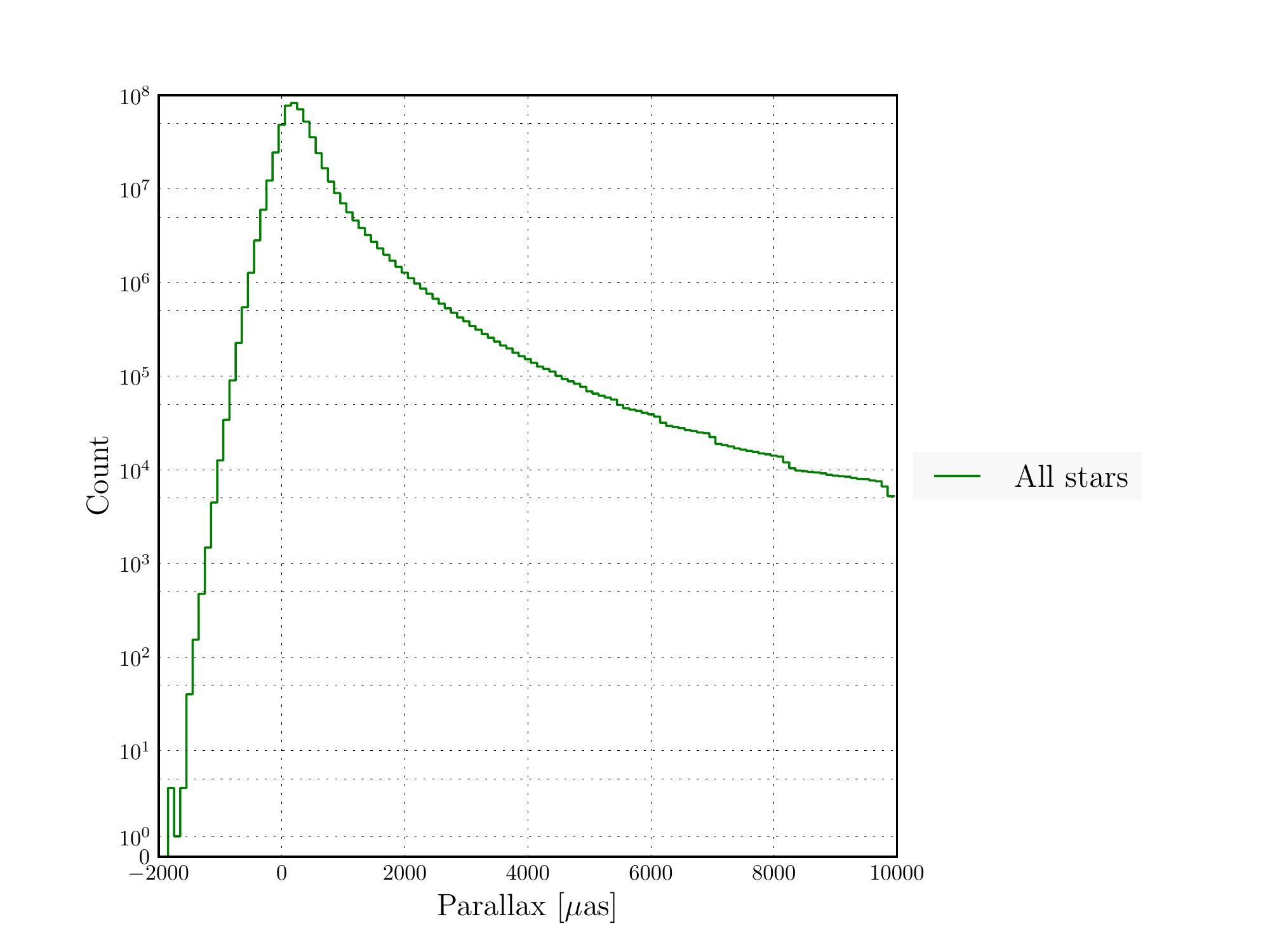} \caption{Histogram of parallax for all single stars. The histogram contains 99.5\% of all data.}
\label{fig:parallax}
\end{minipage}
\quad
\begin{minipage}[b]{0.45\linewidth}
\includegraphics[width=\linewidth]{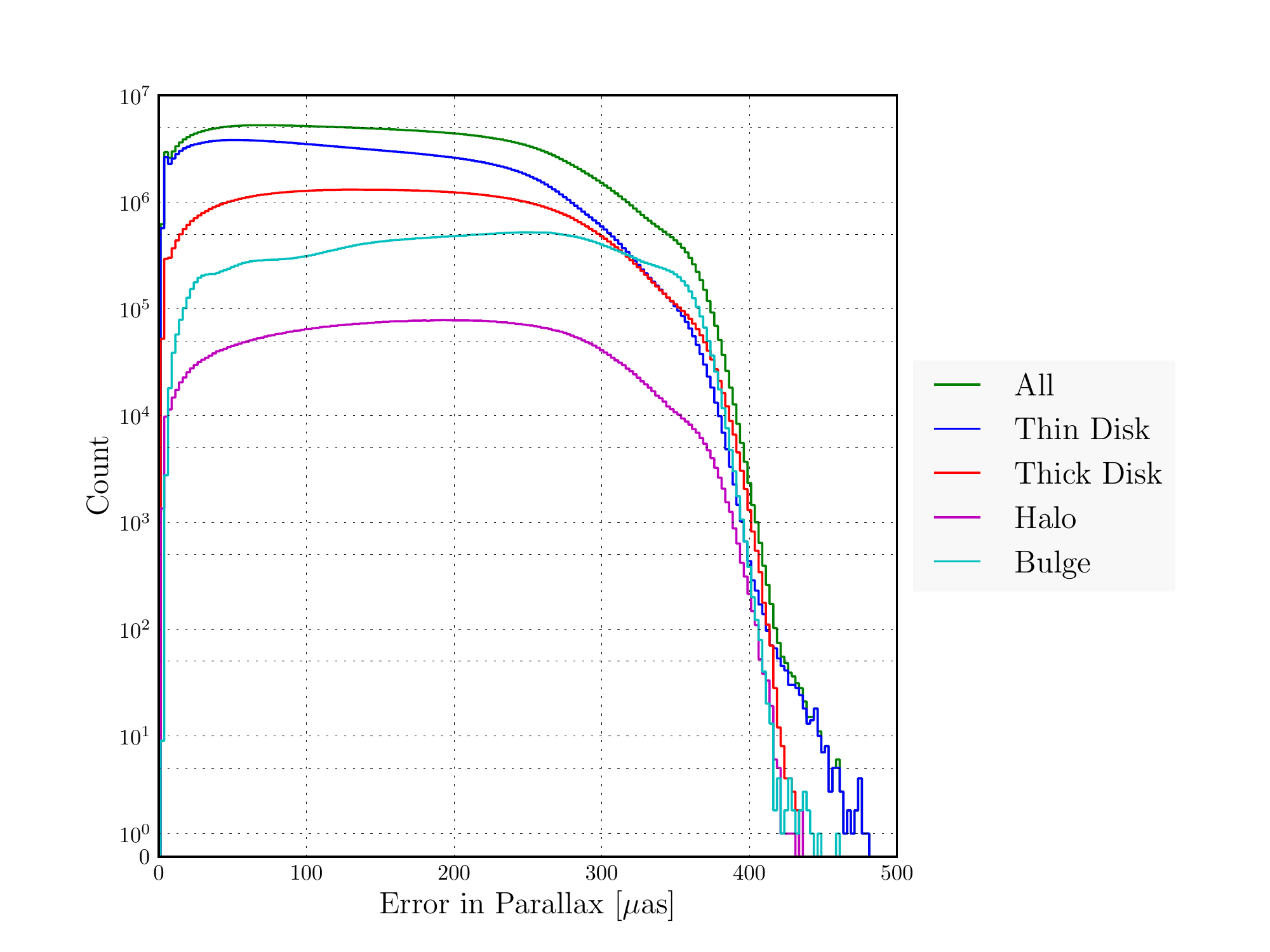} 
\caption{Histogram of end-of-mission parallax error for all single stars, split by stellar population.}
\label{fig:errorpibypop}
\end{minipage}
\end{figure*}

\begin{figure*}[ht]
\centering
\begin{minipage}[b]{0.45\linewidth}
\includegraphics[width=\linewidth]{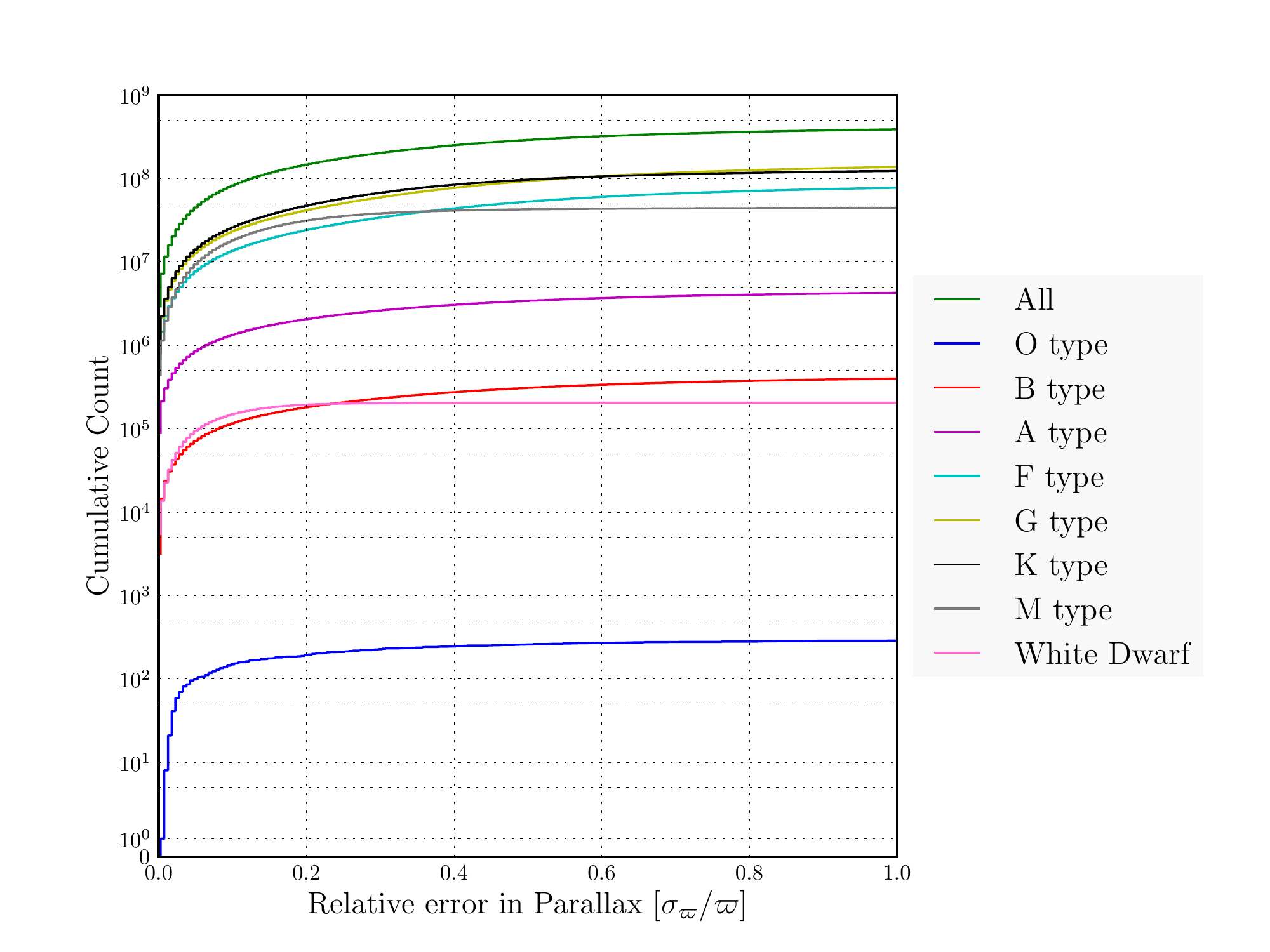} 
\caption{Cumulative histogram of relative parallax error for all single stars, split by spectral type.  The histogram range displays 74\% of all data.}
\label{fig:relerrorpiSpectralType}
\end{minipage}
\quad
\begin{minipage}[b]{0.45\linewidth}
\includegraphics[width=\linewidth]{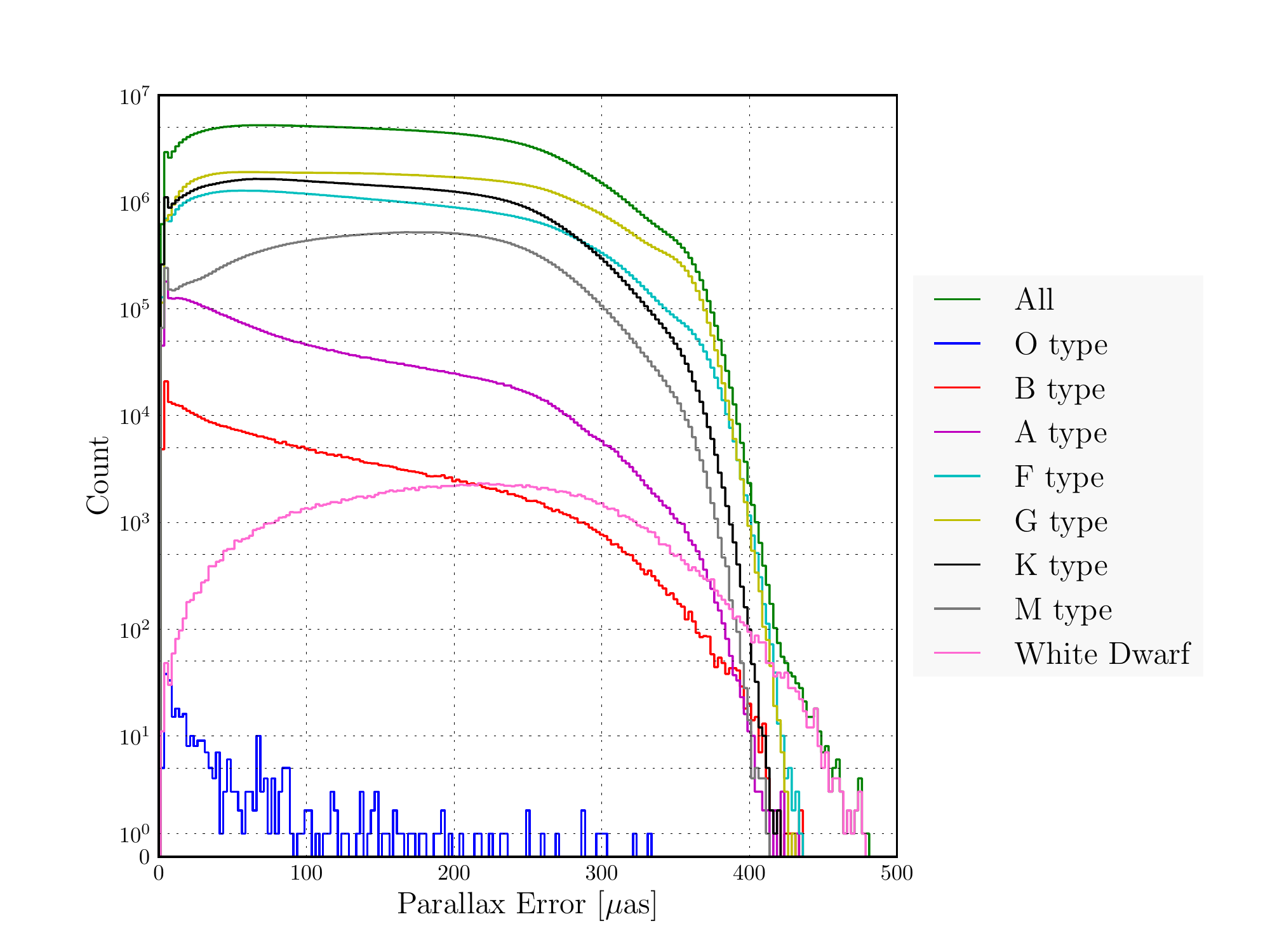} 
\caption{Histogram of end-of-mission parallax error for all single stars, split by spectral type.}
\label{fig:errorpispectraltype}
\end{minipage}
\end{figure*}

\begin{figure*}[ht]
\centering
\begin{minipage}[b]{0.45\linewidth}
\includegraphics[width=\linewidth]{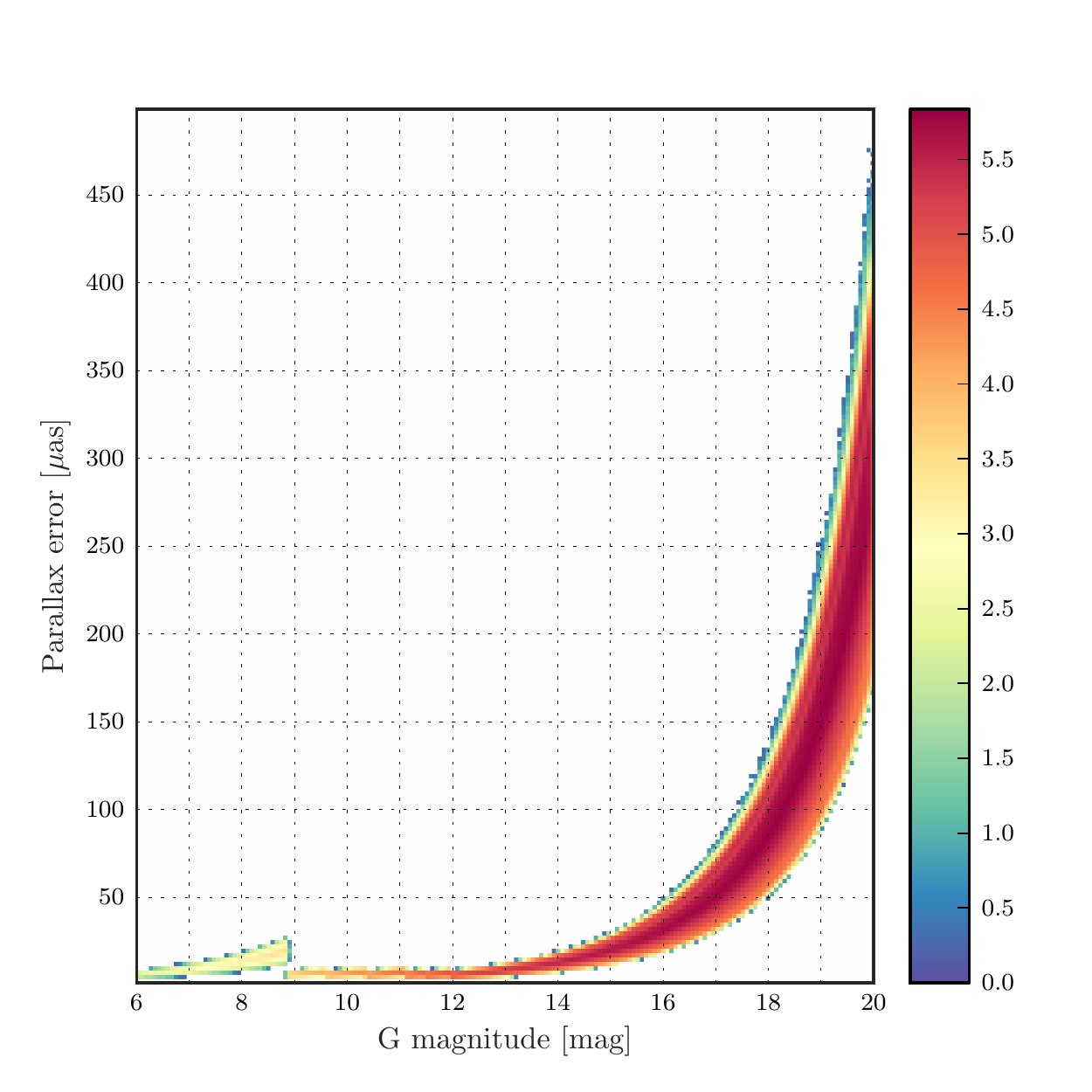} 
\caption{End-of-mission parallax error against $G$ magnitude for all single stars. The colour scale represents the log of density of objects in a bin size of 80 mmag by 2.5 $\mu$as. White area represents zero stars.}
\label{fig:gmagwithparallax}
\end{minipage}
\quad
\begin{minipage}[b]{0.45\linewidth}
\includegraphics[width=\linewidth]{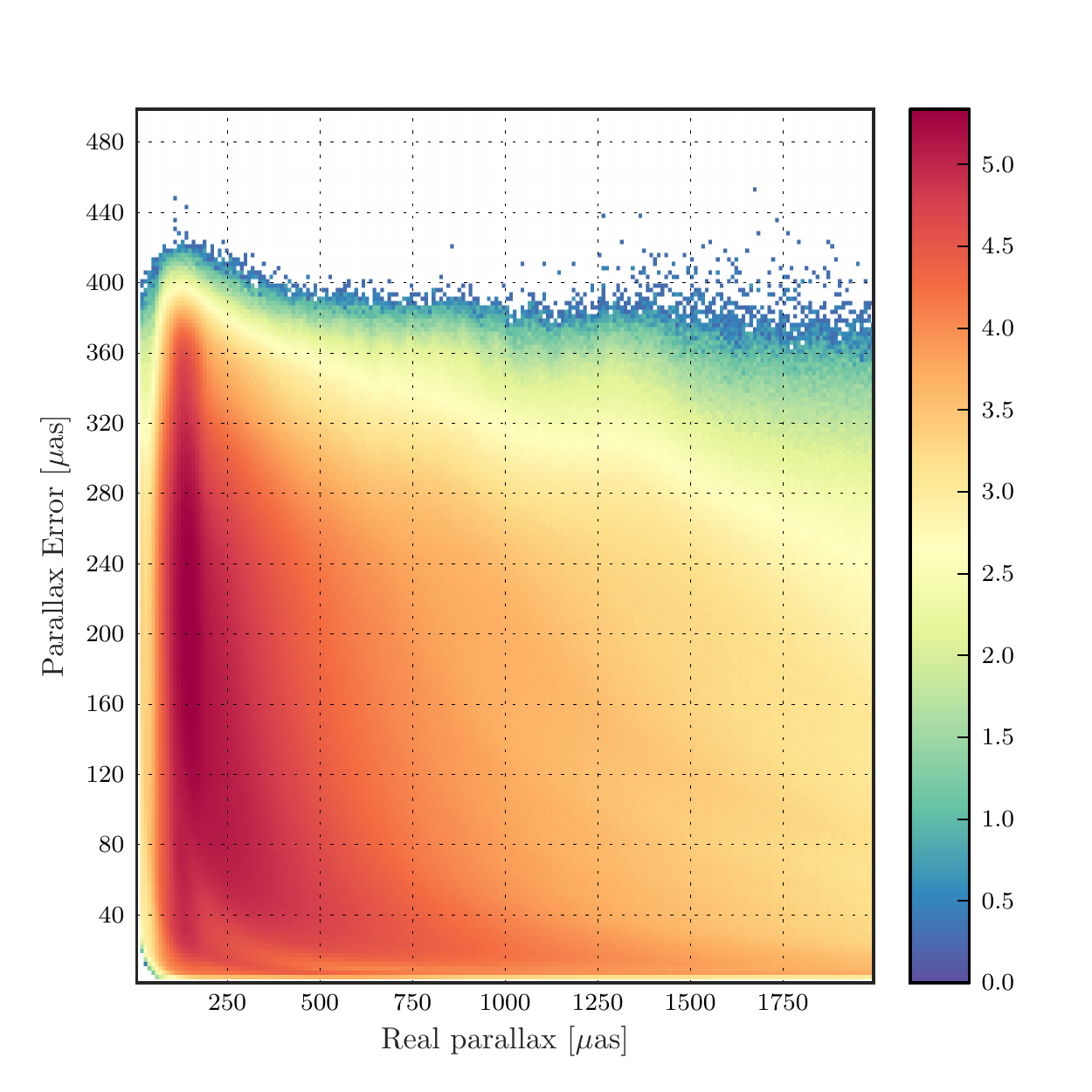} 
\caption{End-of-mission parallax error against parallax for all single stars. The colour scale represents the log of density of objects in a bin size of 10 by 2.5 $\mu$as. White area represents zero stars.}
\label{fig:parallaxwitherror}
\end{minipage}
\end{figure*}

The error in parallax measurements for Gaia depends on the magnitude of the source, the number of observations made, and the true value of the parallax. Figure \ref{fig:parerrormap} shows the mean parallax error over the sky. Its shape clearly follows that of the Gaia scanning law. The red area corresponding to the region of worst precision is due to the bulge population, which suffers from high levels of reddening. The faint ring around the centre of the figure corresponds to the disk of the Galaxy, remembering that the plot is given in equatorial coordinates. The blue areas corresponding to regions of improved mean precision are areas with a higher number of observations. The characteristic shape of this plot is due to the Gaia scanning law. The error in parallax as a function of measured $G$ magnitude is given in Fig. \ref{fig:gmagwithparallax} and as a function of the real parallax in Fig. \ref{fig:parallaxwitherror}.

\begin{table}\centering
    \begin{tabular}{@{}lllll@{}}\toprule 
        
        Spec. type & Total & $\sigma_{\varpi}/ \varpi <  1$ & $\sigma_{\varpi}/\varpi <  0.2$ & $\sigma_{\varpi}/ \varpi <0.05$ \\ \midrule
     O   &  3.3$\times$10$^2$   &   87.5  &  57.5   & 29.2  \\ 
     B   &  3.4$\times$10$^5$   &   74.0  &  33.0   & 12.2  \\ 
     A   &  5.3$\times$10$^6$   &   79.7  &  38.0   & 14.7  \\ 
     F   &  1.2$\times$10$^8$   &   66.2  &  20.1   &  6.1  \\
     G   &  2.0$\times$10$^8$   &   67.4  &  20.0   &  5.7  \\ 
     K   &  1.5$\times$10$^8$   &   82.4  &  30.9   &  8.6  \\ 
     M   &  4.5$\times$10$^7$   &   98.1  &  68.0   & 18.6  \\ 
        \bottomrule
    \end{tabular}
    \caption{Total number of single star for each spectral type, along with the percentage of those that fall below each relative parallax error limit; e.g., 68\% of M-type stars have a relative parallax error better than 20\%.}
    \label{tab:relerrorpiSpectraltype}

\end{table}

\begin{table}\centering
    \begin{tabular}{@{}lllll@{}}\toprule 
        
        Lum. class & Total & $\sigma_{\varpi}/ \varpi <  1$ & $\sigma_{\varpi}/\varpi <  0.2$ & $\sigma_{\varpi}/ \varpi <0.05$ \\ \midrule
Supergiant         & $5.6\times$10$^3$  &  91.5  &  65.9  &  36.8  \\  
Bright giant       & $6.9\times$10$^5$  &  87.1  &  57.9  &  25.1  \\ 
Giant              & $6.6\times$10$^7$  &  67.5  &  21.4  &  7.0   \\ 
Sub-giant          & $7.5\times$10$^7$  &  60.2  &  16.8  &  5.3   \\ 
Main sequence      & $3.8\times$10$^8$  &  78.1  &  30.7  &  8.5   \\ 
White dwarf        & $2.1\times$10$^5$  & 100.0  &  94.3  & 41.9   \\ 
        \bottomrule
    \end{tabular}
    \caption{Total number of single stars for each luminosity class, along with the percentage that fall below each relative parallax error limit.}
        \label{tab:relerrorpiLumClass}

\end{table}

\begin{table}\centering
    \begin{tabular}{@{}lllll@{}}\toprule 
        
        Type & Total & $\sigma_{\varpi}/ \varpi <  1$ & $\sigma_{\varpi}/\varpi <  0.2$ & $\sigma_{\varpi}/ \varpi <0.05$ \\ \midrule
        OII    &   4                 &   100   &  100  &  75   \\ 
        OIII   &   17                &   100   &  65   &  35   \\ 
        OIV    &   26                &   96    &  50   &  31   \\ 
        OV     &   203               &   89    &  57   &  27   \\ 
        BII    &   80                &   91    &  50   &  21   \\ 
        BIII   &   1.5$\times$10$^5$ &   58    &  19   &  8    \\ 
        BIV    &   1.1$\times$10$^5$ &   81    &  42   &  17   \\ 
        BV     &   2.1$\times$10$^5$ &   81    &  38   &  13   \\ 
        AII    &   6.7$\times$10$^3$ &   79    &  42   &  19   \\ 
        AIII   &   9.5$\times$10$^5$ &   76    &  37   &  16   \\ 
        AIV    &   1.4$\times$10$^6$ &   80    &  40   &  16   \\ 
        AV     &   2.7$\times$10$^6$ &   81    &  37   &  14   \\ 
        FII    &   2.0$\times$10$^3$ &   81    &  46   &  22   \\ 
        FIII   &   1.7$\times$10$^6$ &   76    &  33   &  12   \\ 
        FIV    &   3.6$\times$10$^7$ &   67    &  22   &  7    \\ 
        FV     &   7.9$\times$10$^7$ &   66    &  19   &  5    \\ 
        GII    &   1.6$\times$10$^5$ &   81    &  43   &  20   \\ 
        GIII   &   2.0$\times$10$^7$ &   61    &  17   &  5    \\ 
        GIV    &   3.7$\times$10$^7$ &   53    &  11   &  3    \\ 
        GV     &   1.5$\times$10$^8$ &   72    &  23   &  6    \\ 
        KII    &   2.5$\times$10$^5$ &   81    &  43   &  19   \\ 
        KIII   &   4.0$\times$10$^7$ &   69    &  21   &  7    \\ 
        KV     &   1.1$\times$10$^8$ &   87    &  34   &  9    \\ 
        MII    &   1.1$\times$10$^4$ &   89    &  59   &  32   \\ 
        MIII   &   2.2$\times$10$^6$ &   85    &  46   &  16   \\ 
        MV     &   4.2$\times$10$^7$ &   99    &  70   &  19   \\
        WD        & $2.1\times$10$^5$  & 100   &  94   &  42   \\ 
        \bottomrule
    \end{tabular}
    \caption{Total number of single stars for each stellar classification, along with the percentage that fall below each relative parallax error limit.}
    \label{tab:relerrorpiByAll}

\end{table}

\subsubsection{Position}

Gaia will be capable of measuring the position of each observed star at an unprecedented accuracy, producing the most precise full sky position catalogue to date. 

\begin{figure*}[ht]
\centering
\begin{minipage}[b]{0.45\linewidth}
\includegraphics[width=\linewidth]{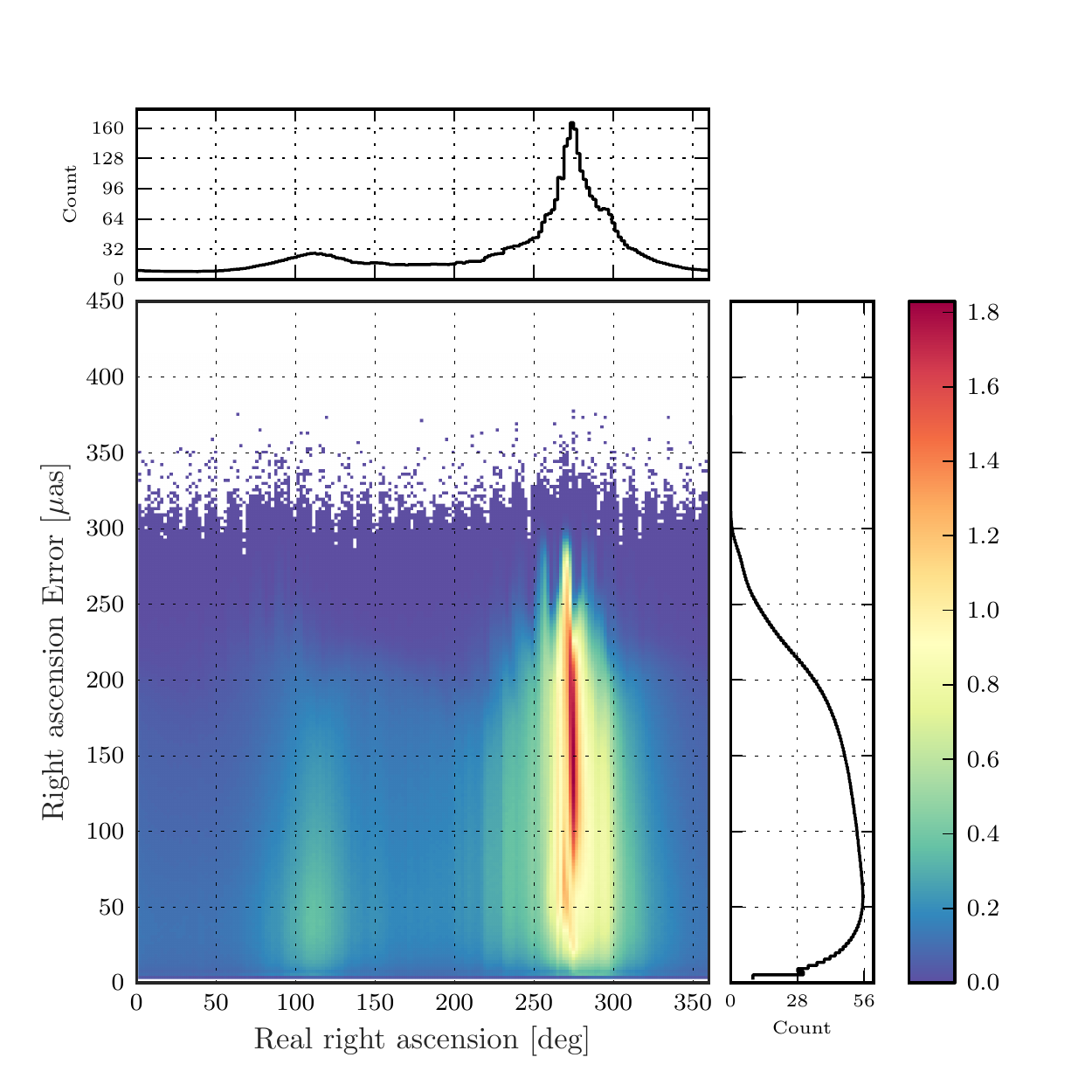} \caption{Right ascension error against real right ascension. The colour scale is linear, with a factor of $10^5$. Histograms are computed for both right ascension and right ascension error. The colour scale represents log density of objects in a bin size of 2 degrees by 7.5 $\mu$as. White area represents zero stars.}
\label{fig:RAvsError}
\end{minipage}
\quad
\begin{minipage}[b]{0.45\linewidth}
\includegraphics[width=\linewidth]{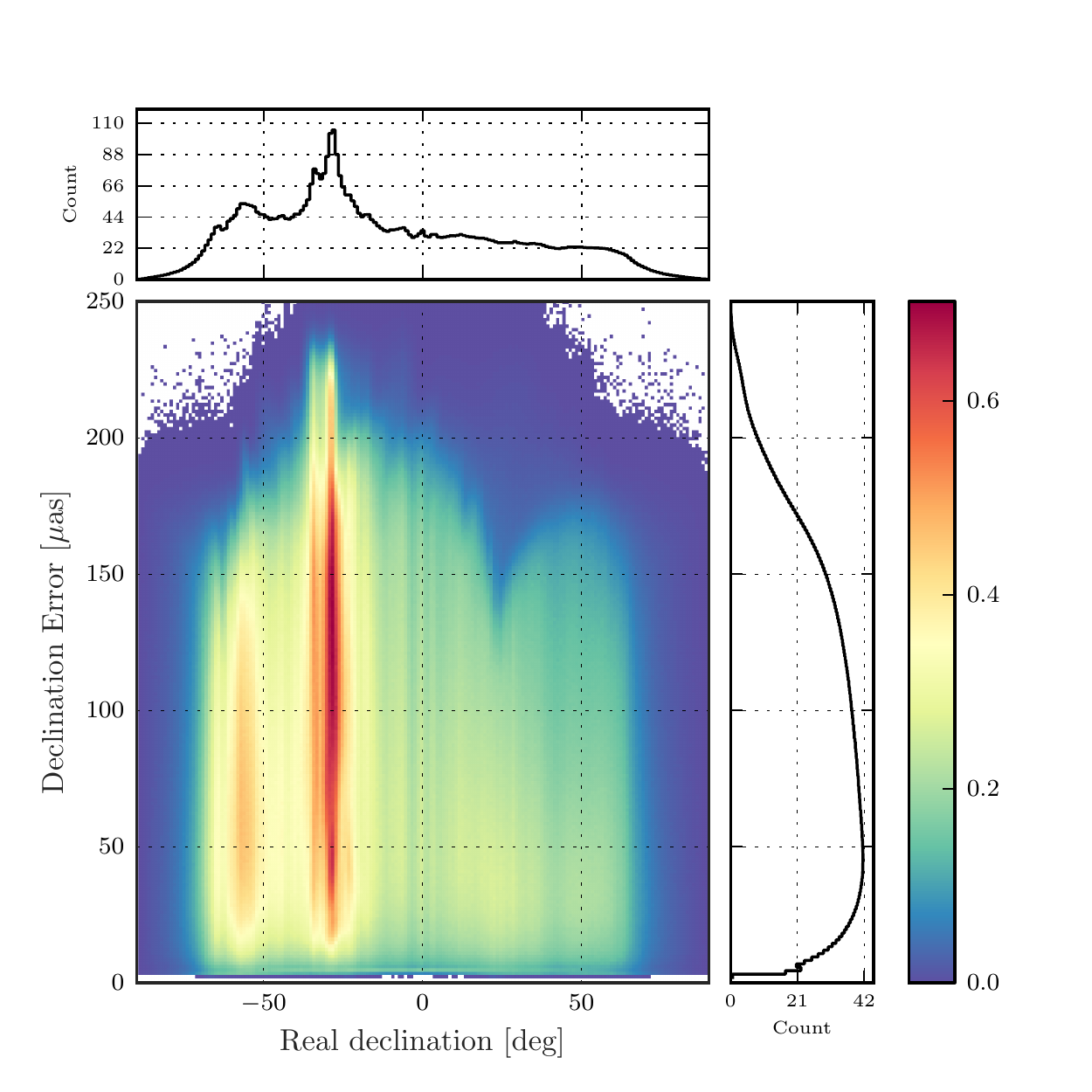} 
\caption{Declination error against real declination. The colour scale is linear, with a factor of $10^5$. Histograms are computed for both declination and declination error. The colour scale represents log density of objects in a bin size of 1 degrees by 5 $\mu$as. White area represents zero stars.}
\label{fig:DECvsError}
\end{minipage}
\end{figure*}

The mean error is 90 $\mu$as for right ascension and 103 $\mu$as for declination. The distribution of error in right ascension and declination as a function of the true value, along with a histogram of the error, are given in Figs. \ref{fig:RAvsError} and \ref{fig:DECvsError}. The overdensities are due to the bulge of the Galaxy.

\subsubsection{Proper motion and radial velocity}

\begin{figure}
\includegraphics[width=\linewidth]{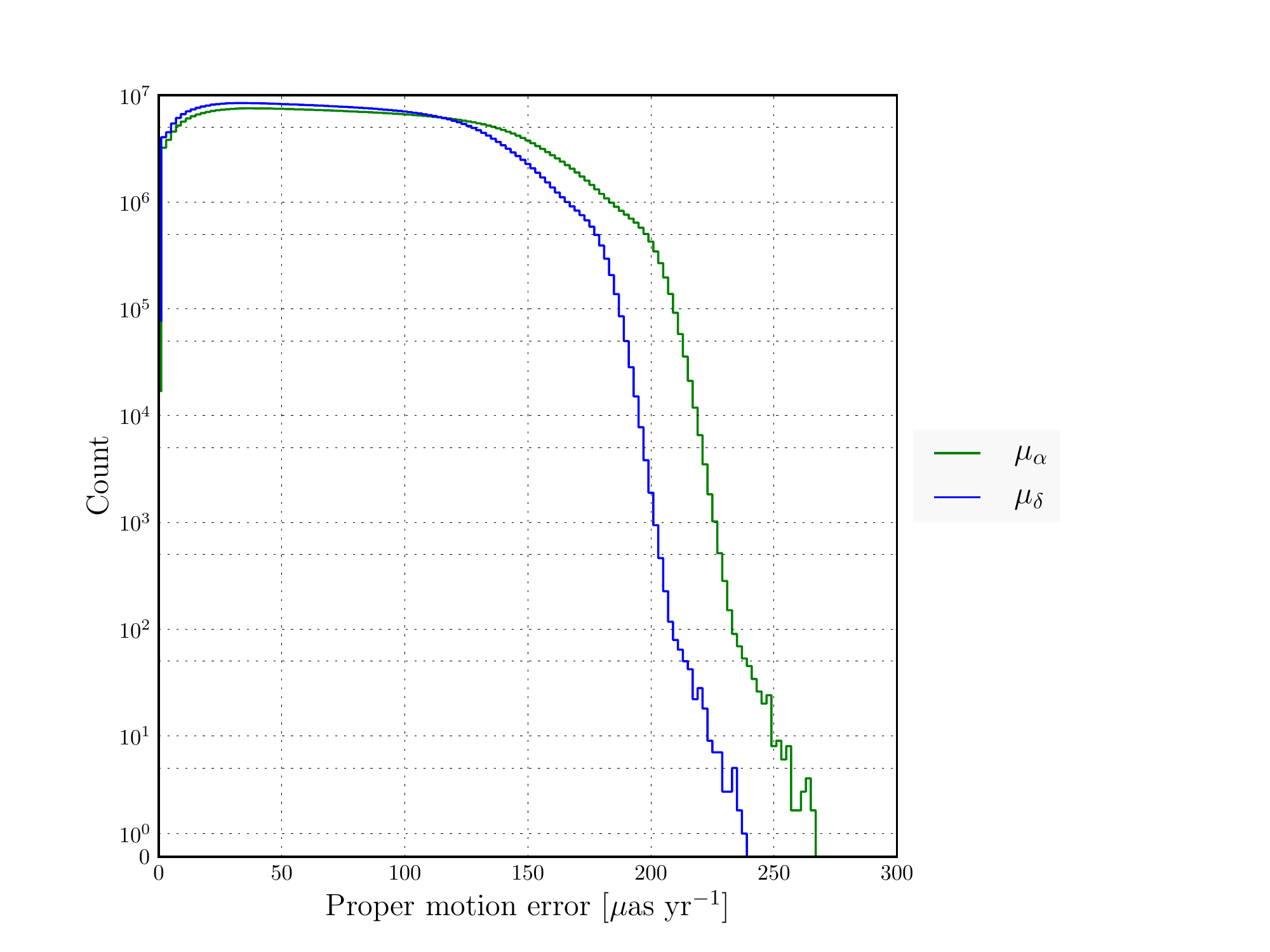} 
\caption{Error in proper motion for alpha and delta for all single stars.}
\label{fig:propermotionerror}
\end{figure}

In addition to parallax measurements, Gaia will also measure proper motions for all stars it detects. The proper motion in right ascension and declination is labelled as $\mu_\alpha$ and $\mu_\delta$, respectively. The mean error in $\mu_\alpha$ is 81.7 $\mu$as$\cdot$yr$^{-1}$, and in $\mu_\delta$ is 72.9 $\mu$as$\cdot$yr$^{-1}$. The distribution of errors in both components of proper motion is given in Fig. \ref{fig:propermotionerror}.

The radial velocity is measured by the on-board radial velocity spectrometer. This instrument is only sensitive to stars down to $G_{RVS}=17$ magnitude. We assume an upper limit on the error in radial velocity of 20 km$\cdot$s$^{-1}$, and assume that stars with a precision worse than this will not be given any radial velocity information. 

\begin{figure*}[ht]
\centering
\begin{minipage}[b]{0.45\linewidth}
\includegraphics[width=\linewidth]{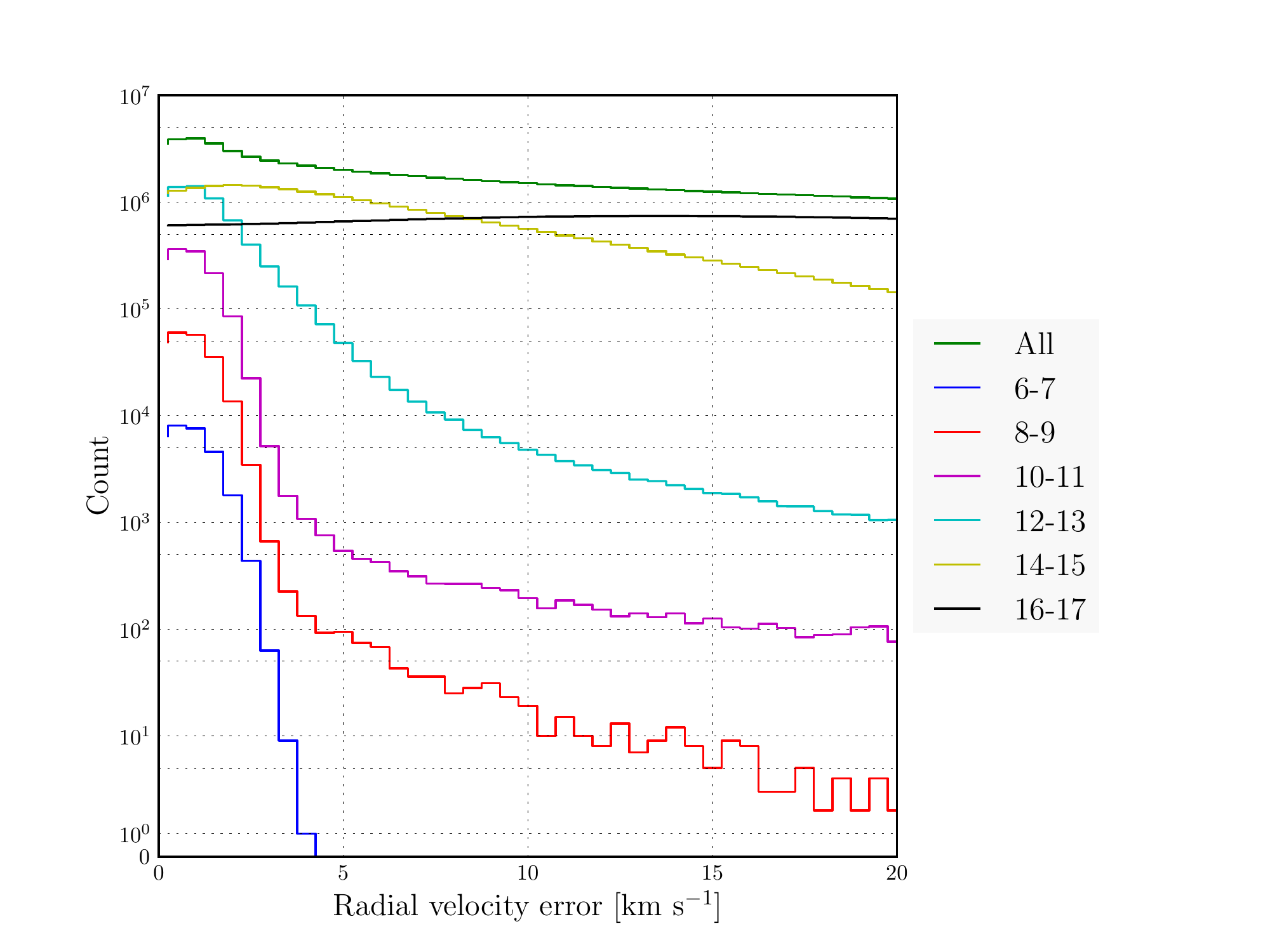} 
\caption{Histogram of radial velocity error split by $G$ magnitude range. The histogram contains 100\% of all data that have radial velocity information.}
\label{fig:radialvelocityerrorM}
\end{minipage}
\quad
\begin{minipage}[b]{0.45\linewidth}
\includegraphics[width=\linewidth]{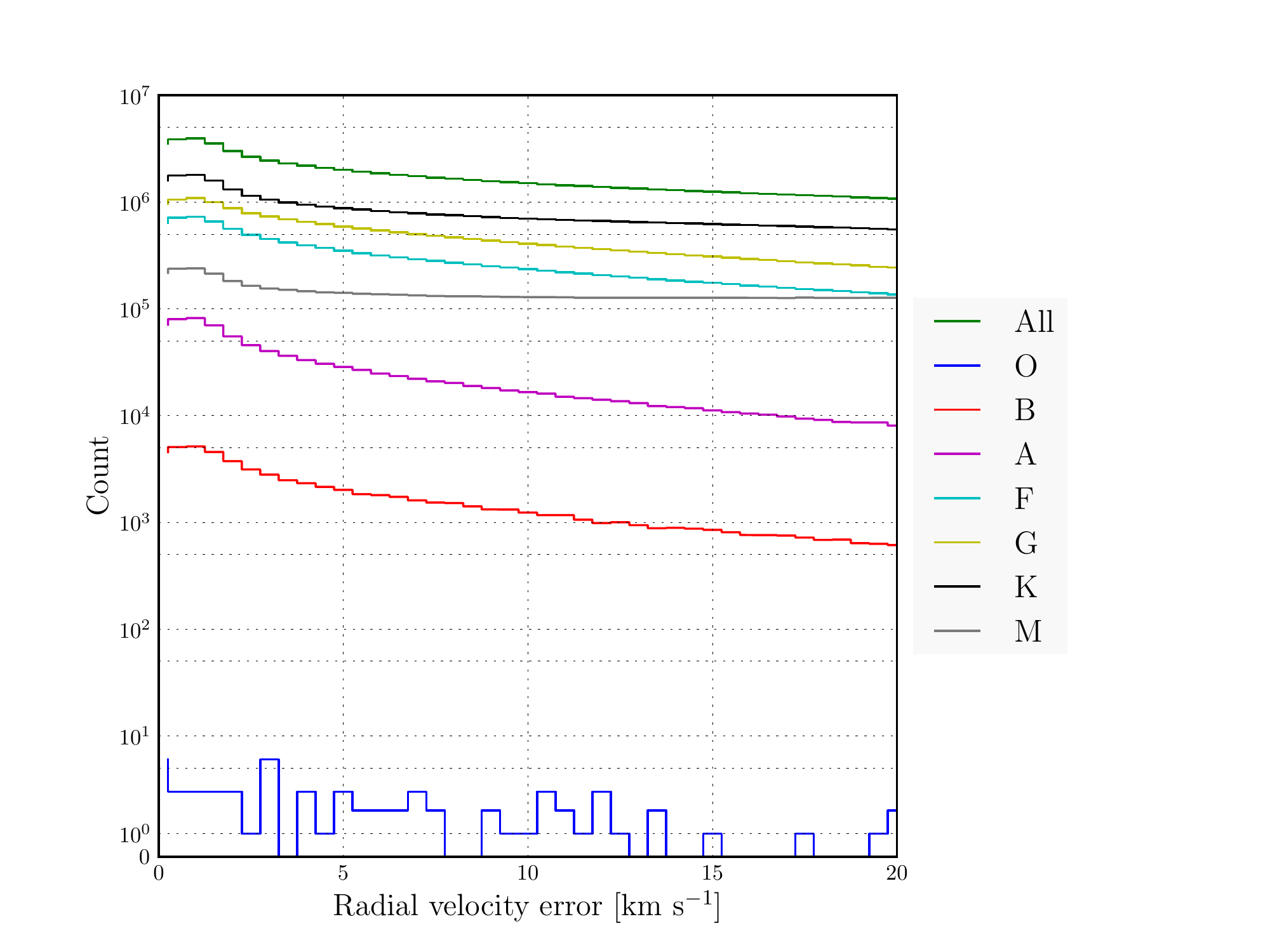} 
\caption{Histogram of radial velocity error split by spectral type. The histogram contains 100\% of all data that have radial velocity information.}
\label{fig:radialvelocityerrorST}
\end{minipage}
\end{figure*}

Of the 523 million measured individual Milky Way stars, 74 million have a radial velocity measurement. The mean error in the radial velocity measurement is 8.0 km$\cdot$s$^{-1}$. The distribution of radial velocity error is given for each $G$ magnitude in Fig. \ref{fig:radialvelocityerrorM}, and in Fig. \ref{fig:radialvelocityerrorST} split by spectral type. The radial velocity error is given as a function of $G_{RVS}$ magnitude in Fig. \ref{fig:RVS}

\begin{figure*}[ht]
\centering
\subfigure{%
\includegraphics[width=0.45\linewidth]{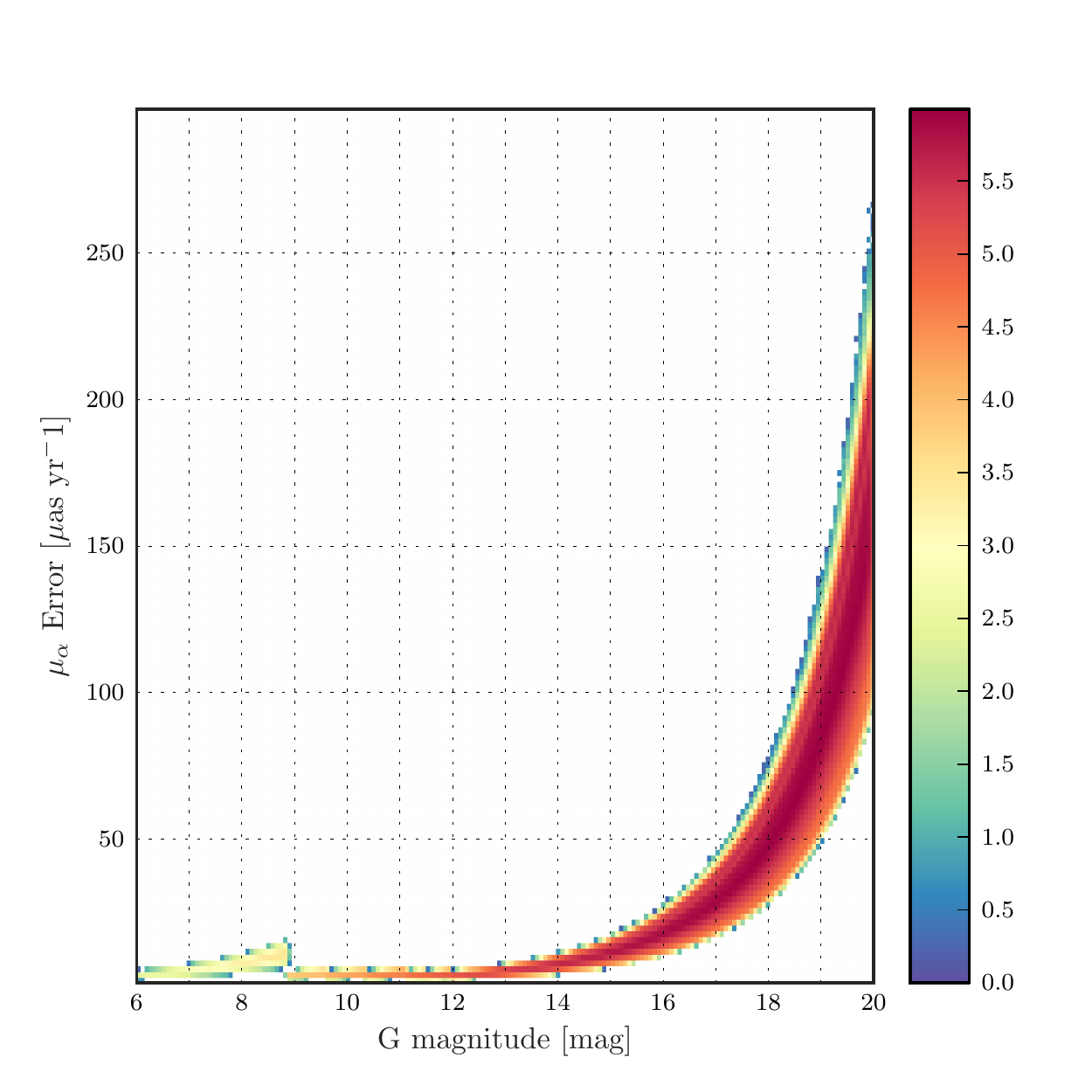} }
\quad
\subfigure{%
\includegraphics[width=0.45\linewidth]{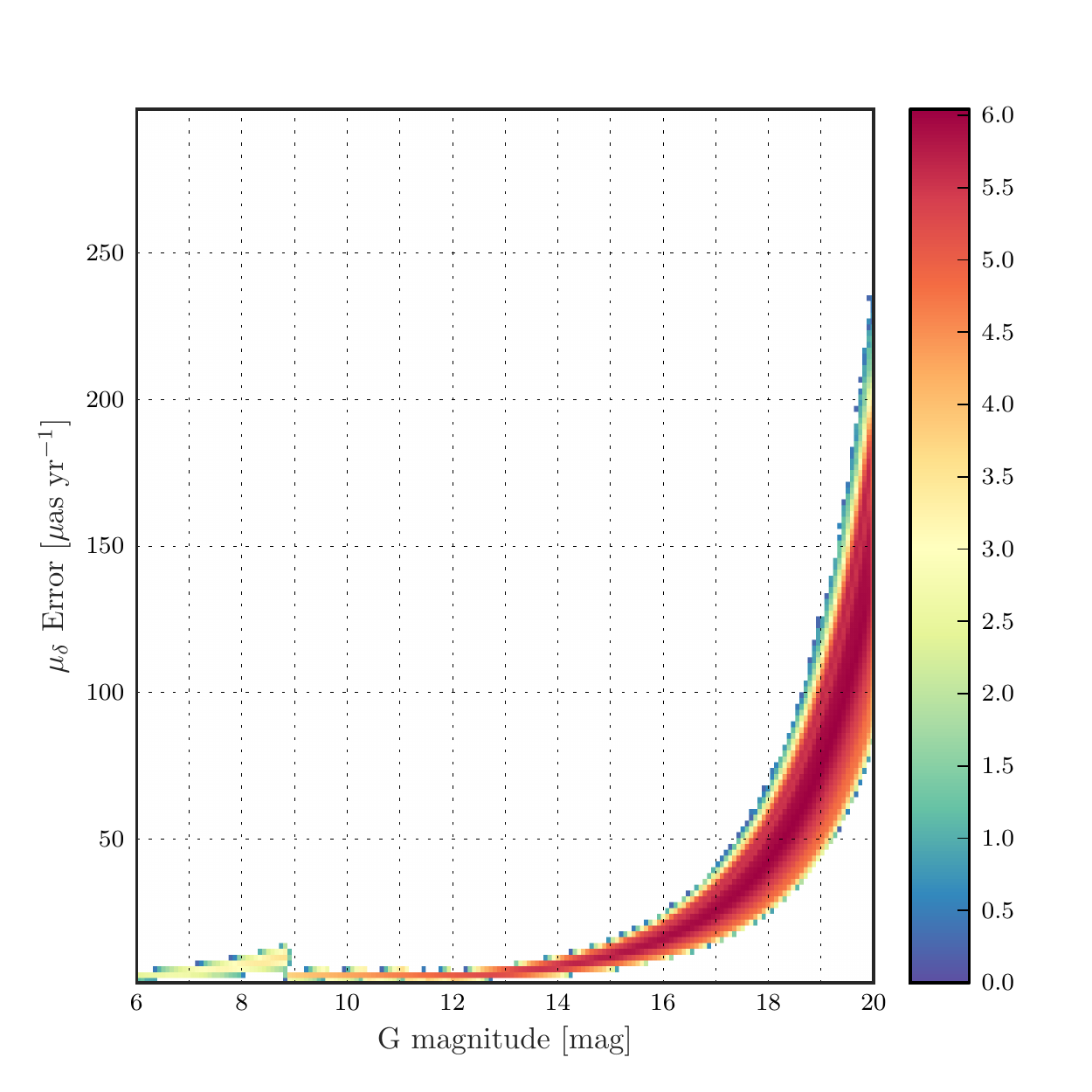} }
    \caption{2D histograms showing error in proper motion against $G$ magnitude. The colour scale represents log density of objects in a bin size of 80 mmag by 2 $\mu$as$\cdot$yr$^{-1}$. Left is proper motion in right ascension, and right is proper motion in declination. White area represents zero stars.}
    \label{fig:muAlphaAndDeltaWithG}
\end{figure*}

\begin{figure}
\centering
\includegraphics[width=\linewidth]{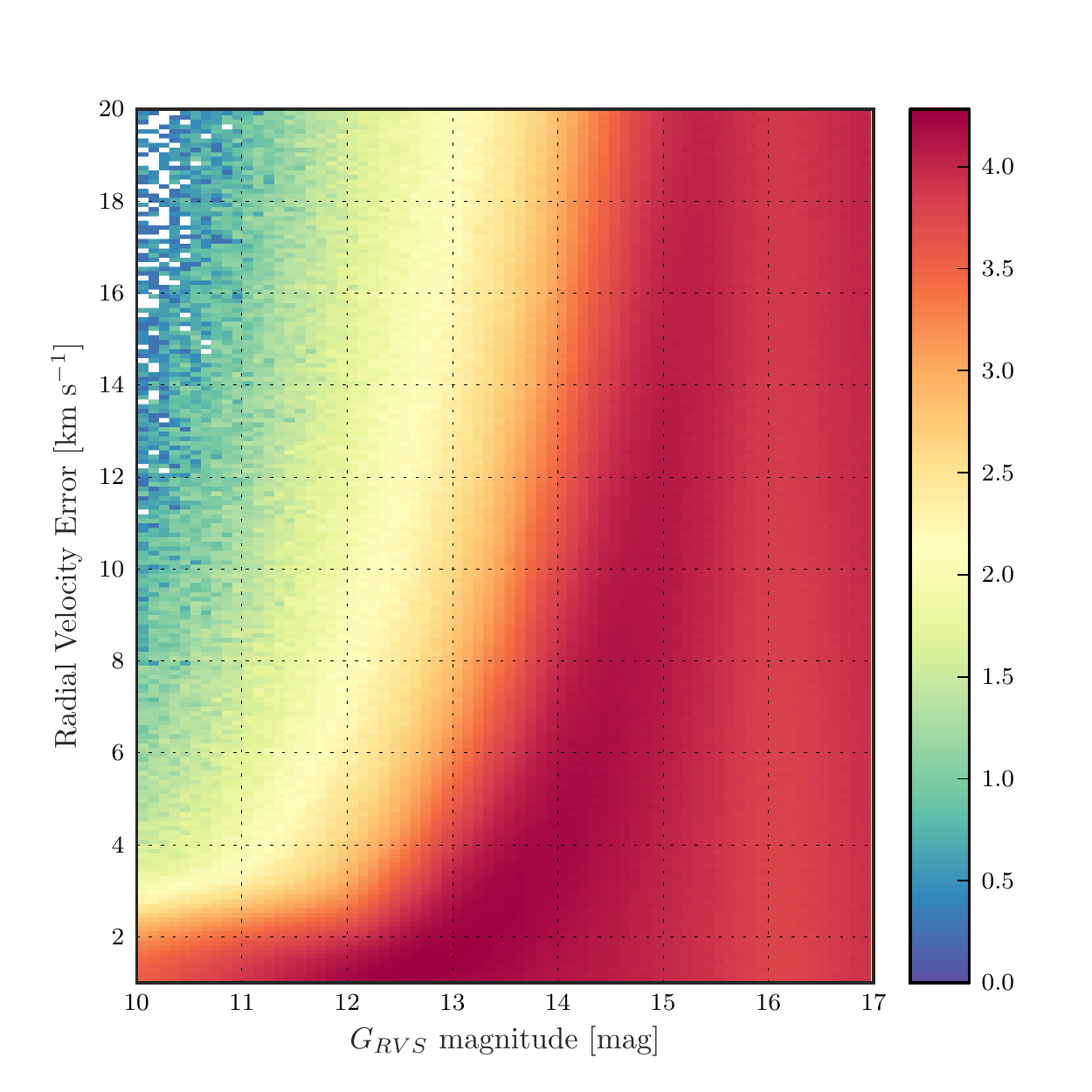} 
    \caption{End-of-mission error in radial velocity against $G_{RVS}$ magnitude. The colour scale represents log density in a bin size of 50 mmag by 1 km$\cdot$s$^{-1}$. White area represents zero stars.}
    \label{fig:RVS}
\end{figure}

\subsubsection{Photometry}

\begin{figure*}
\begin{center}
\includegraphics[width=\linewidth]{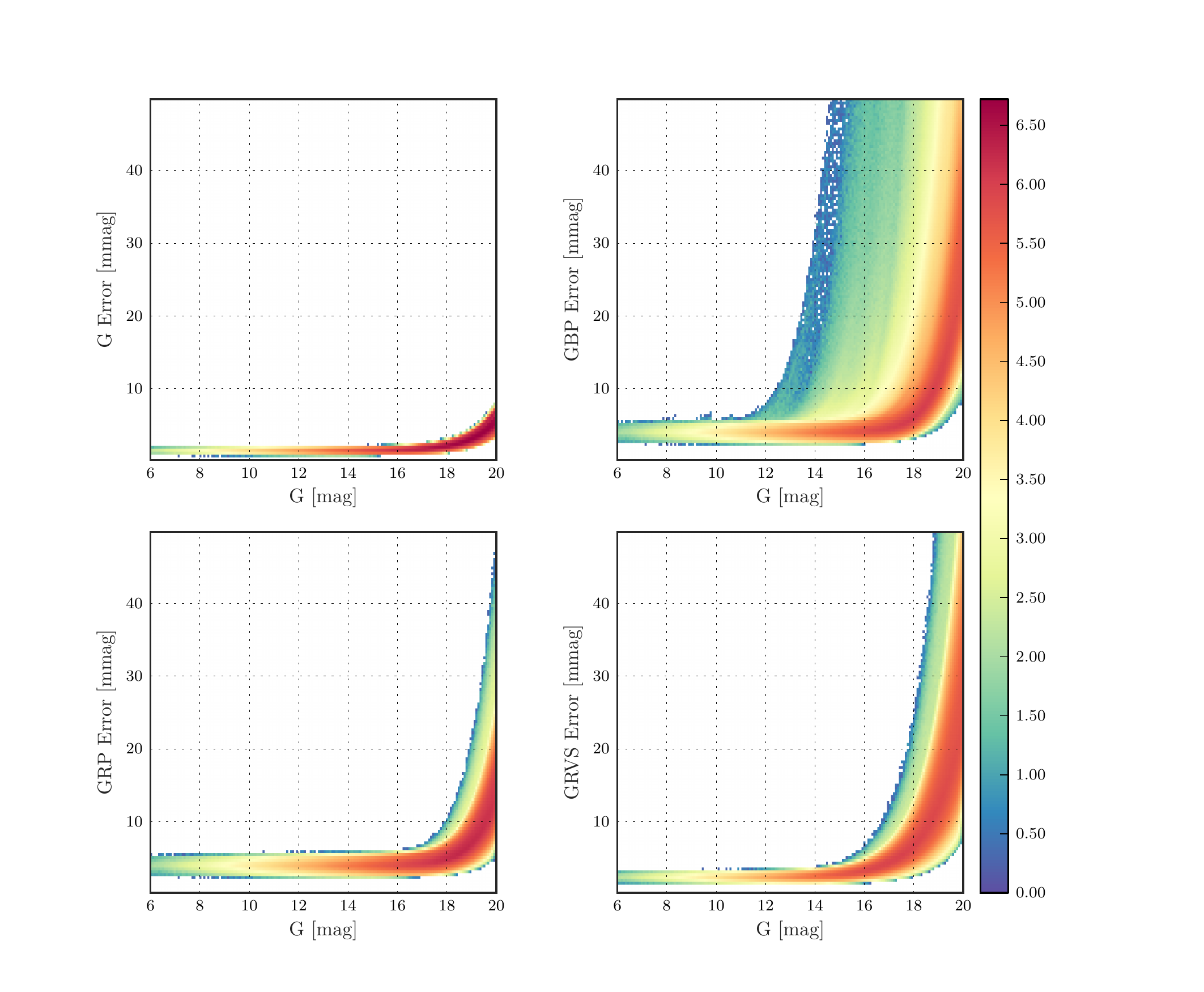} 
\caption{End-of-mission errors in photometry as a function of $G$ magnitude. The colour scale represents log density of objects in a bin size of 80 mmag by 0.4 mmag. Top left, $G$ magnitude; top right, $G_{BP}$; bottom left, $G_{RP}$; bottom right, $G_{RVS}$. White area represents zero stars. }
\label{fig:photometryerrors}
\end{center}    
\end{figure*}

 The end-of-mission error in each measurement as a function of $G$ magnitude is given in Fig. \ref{fig:photometryerrors}. 

Gaia will produce low-resolution spectra, in addition to measuring the magnitude of each source in the Gaia bands $G$, $G_{BP}$, $G_{RP}$, and $G_{RVS}$. Whilst GOG is capable of simulating these spectra, they have not been included in the present simulations owing to the long computation time and the large storage space requirement of a catalogue of spectra for one billion sources. 

\begin{figure}
\begin{center}
\includegraphics[width=\linewidth]{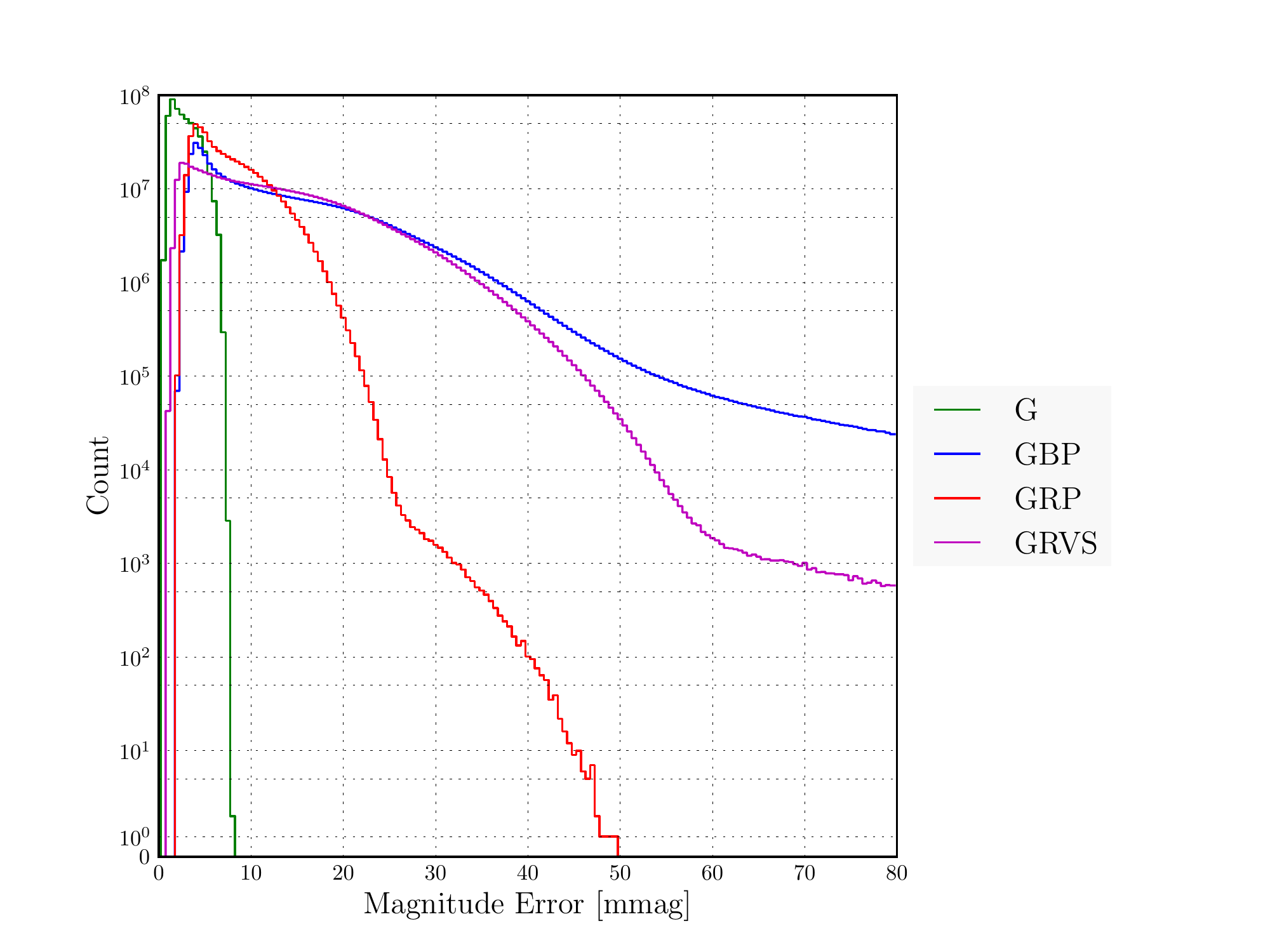} 
\caption{Histogram of error in $G$,  $G_{RVS}$, $G_{RP}$, and $G_{BP}$ for all single stars.}
\label{fig:photometryHist}
\end{center}    
\end{figure}

Figure \ref{fig:photometryHist} shows the distribution in the error of each photometric measurement. As can be seen in this figure, the error in $G$ is much lower than for the other instruments, and for all stars it is less than 8 mmag. The mean error in $G$ is 3.0 mmag. The mean error in $G_{BP}$ and $G_{RP}$ is 14.6 mmag and  7.7 mmag, respectively. The mean error in $G_{RVS}$ is 13.2 mmag, although it must be remembered that the radial velocity spectroscopy instrument is limited to brighter than $G_{RVS}=17$.

\begin{figure*}[H]
\begin{center}
\includegraphics[width=0.45\linewidth]{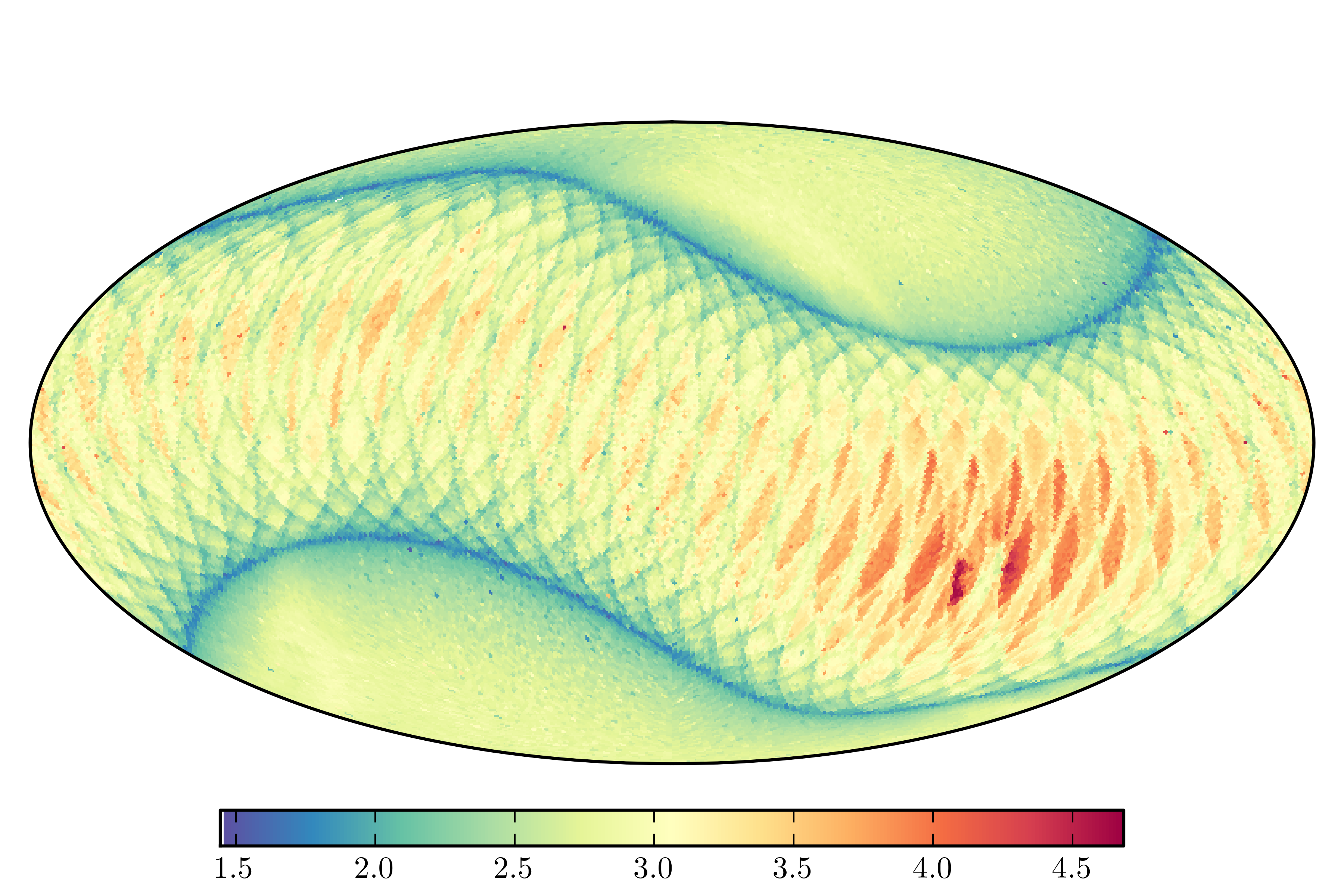} 
\includegraphics[width=0.45\linewidth]{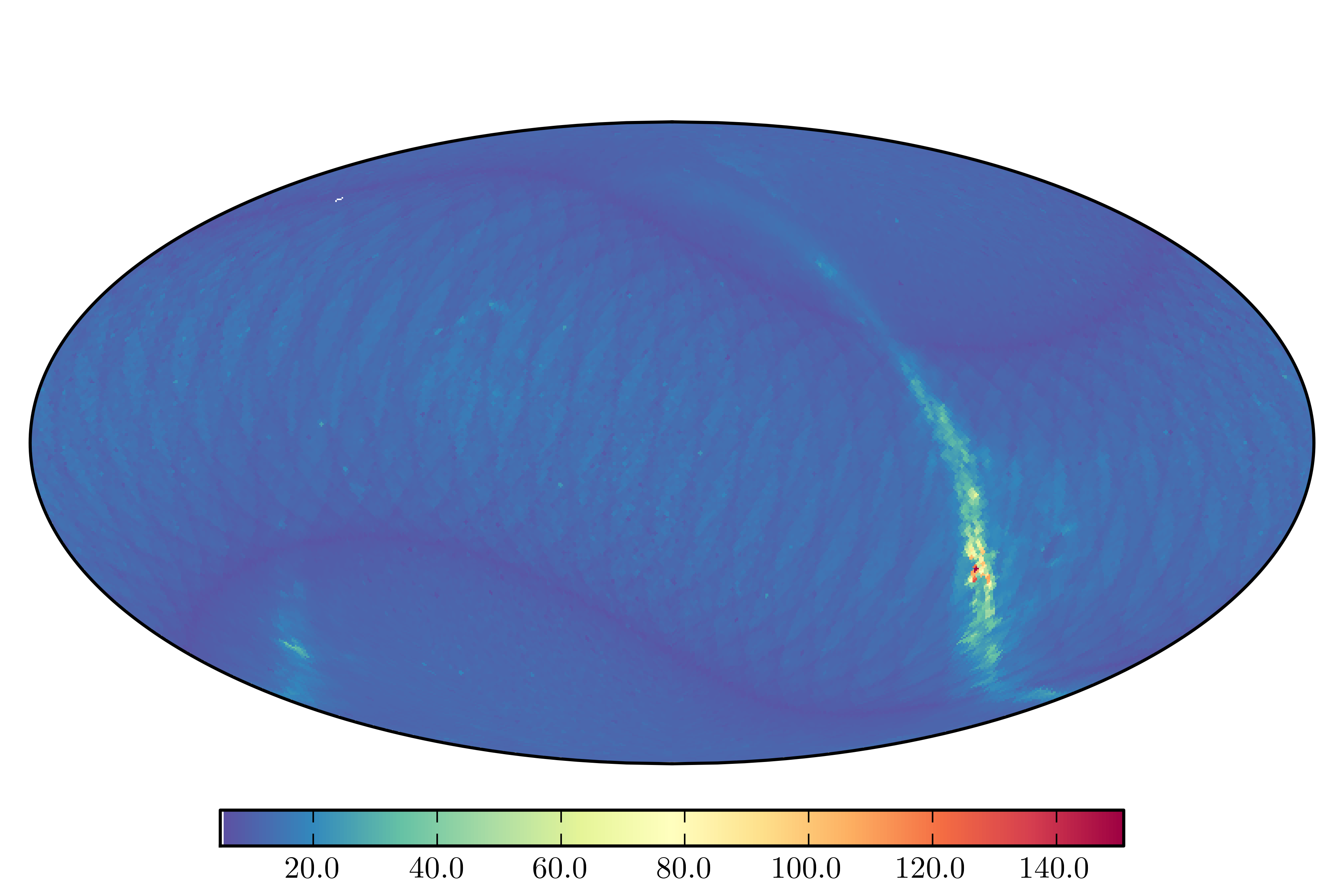} 
\includegraphics[width=0.45\linewidth]{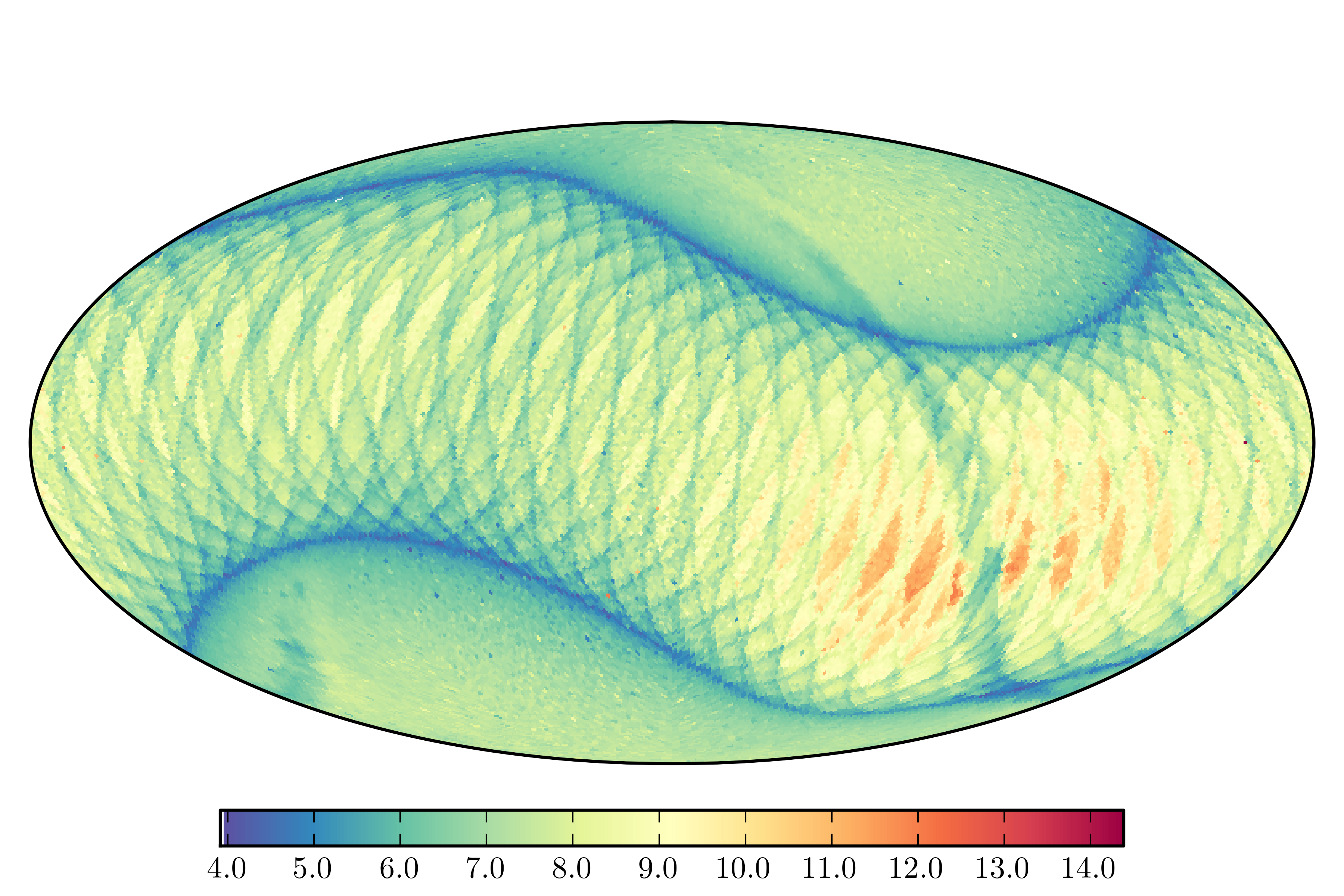} 
\includegraphics[width=0.45\linewidth]{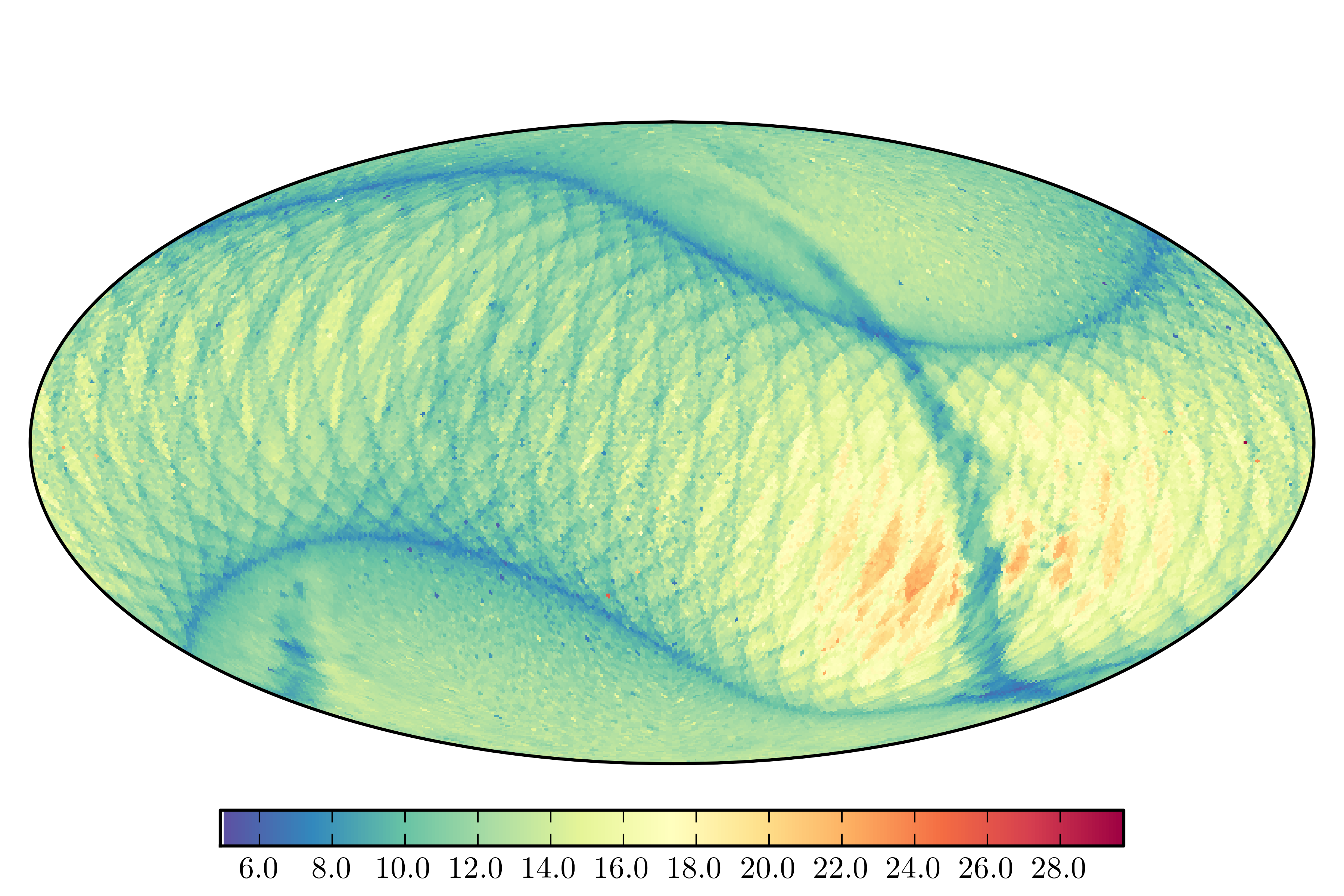} 
\caption{HealPixMap in equatorial coordinates of the mean error in: \emph{Top left:} $G$; \emph{top right:} $G_{BP}$; \emph{lower left:} $G_{RP}$;\emph{ lower right:} $G_{RVS}$. The colour scale gives the mean photometric error in mmag. The colour scales are different due to differences in the maximum mean magnitude.}
\label{fig:photometryMaps}
\end{center}    
\end{figure*}

Figure \ref{fig:photometryMaps} shows the mean photometric error as a function of position on the sky for the four Gaia photometric passbands. The structure seen in all four maps is derived from the Gaia scanning law.

It is interesting to point out the ring in the four plots of Fig. \ref{fig:photometryMaps} caused by the disk of the Galaxy. Owing to significant levels of interstellar dust in the disk of the Galaxy, visible objects are generally much redder. This reddening causes objects to lose flux at the bluer end of the spectrum, making them appear fainter to the $G_{BP}$ photometer. Therefore the plane of the Galaxy can be seen as an \emph{increase} in the mean photometric error in the $G_{BP}$ error map.

Conversely, the disk of the Galaxy shows as a ring of \emph{decreased} mean photometric error in the $G_{RP}$ and $G_{RVS}$ maps, since the sensitivity of their spectra is skewed more towards the redder end of the spectrum. It is important to note, however, that the effect of crowding on photometry is not accounted for in GOG.

\subsection{Variables}\label{sec:Variables}

Gaia will be continuously imaging the sky over its full five-year mission, and each individual object will be observed 70 times on average. The scanning law means that the time between repeated observations varies, and Gaia will be incredibly useful for detecting many types of variable stars. GOG produces a total of 10.8 million single variable objects. This number comes from the UM \citep{robin2011} and assumes 100\% variability detection. The exact detection rates and the classification accuracy for each variability type are still unknown. In fact, the numbers of variable objects in the catalogue is expected to be higher than 10.8 million because some variable star types have not yet been implemented (see \citealt{robin2011} for a more detailed description).

\begin{figure}
\begin{center}
\includegraphics[width=\linewidth]{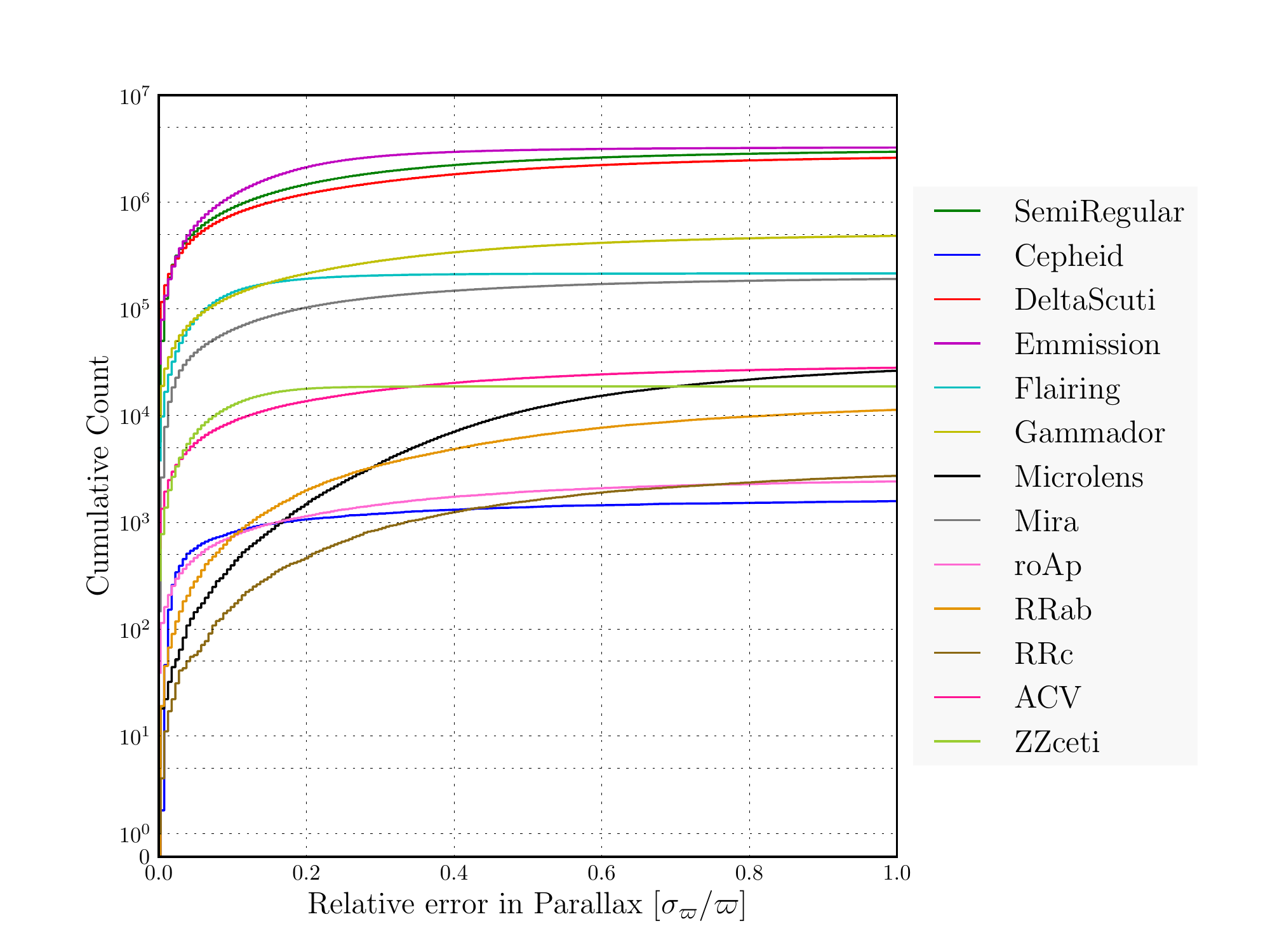} 
\caption{Cumulative histogram of the relative parallax error for all single stars, split by variability type.  The histogram range displays 85\% of all data.}
\label{fig:relerrorpiVar}
\end{center}    
\end{figure}

The distribution of relative parallax error is given for each type of variable star in Fig. \ref{fig:relerrorpiVar}. The numbers of each type of variable produced by GOG are given in Table \ref{tab:variables}, along with the number of each type that falls below each relative parallax error limit.

In general, the numbers of variables presented in this paper are lower than in \cite{robin2011} by a factor of two or three. This is expected, because in the present paper we are excluding all variables that are part of binary or multiple systems, and presenting the number of single variable stars alone.

However, the number of emission variables is higher in the present paper. This is due to implementation of new types of emission stars: Oe, Ae, dMe, and WR stars. These are now included as emission variables but were not simulated in \cite{robin2011}. Additionally, the number of Mira variable stars is higher in the present paper. This is from an implementation error in the version of the UM used in \cite{robin2011}, which has been fixed in the version used in the present paper.

\begin{table*}\centering
    \begin{tabular}{@{}llllllll@{}}\toprule 
       
        Variability type & Total  & $\sigma_{\varpi}/\varpi < 5$  & $\sigma_{\varpi}/\varpi < 1$  & $\sigma_{\varpi}/\varpi < 0.5 $  & $\sigma_{\varpi}/\varpi < 0.2$ & $\sigma_{\varpi}/\varpi < 0.05$  & $\sigma_{\varpi}/\varpi < 0.01$ \\ \midrule
Non-variables & 5.1  $\times10^8$   &  88& 74& 55& 27&7.7&1.4 \\
Emission    &   3.3  $\times10^6$   &  99& 97& 92& 62& 16&2.4 \\
Flaring     &   2.1  $\times10^5$   &  99& 99& 98& 88& 33& 4.6 \\
$\delta$ Scuti  &   3.3  $\times10^6$   &  90& 78& 61& 35& 13& 3.5 \\
Semiregular &   3.6  $\times10^6$   &  92& 82& 68& 40& 14& 1.4 \\
Gammador    &   6.0  $\times10^5$   &  91& 80& 63& 35& 13& 3.2 \\
RR Lyrae AB-type        &   2.4  $\times10^4$   &  67& 45& 25& 7.9&1.0&0.1 \\
Mira        &   2.3  $\times10^5$   &  92& 83& 70& 44& 16& 1.2  \\
ZZ Ceti      &   1.9  $\times10^4$   &  100&100&99& 94& 33& 4.1  \\
ACV         &   3.5  $\times10^4$   &  91& 80& 64& 39& 15& 3.8 \\
RR Lyrae C-type         &   5.6  $\times10^3$   &  68& 45& 25& 7.8&0.9&0.1 \\
$\rho$ Ap        &   3.0  $\times10^3$   &  92& 82& 65& 38& 14& 3.8 \\
Cepheid     &   1.8  $\times10^3$   &  95& 88& 78& 59& 30& 0.1 \\
    \bottomrule
    \end{tabular}
        \caption{Total number of single stars of each variability type, and the percentage of each that falls below each relative parallax error limit: 500\%, 100\%, 50\%, 20\%, 5\%, and 1\%. }
\label{tab:variables}

\end{table*}

\subsubsection{Cepheids and RR-Lyrae}

\begin{figure*}[ht]
\centering
\begin{minipage}[b]{0.45\linewidth}
\includegraphics[width=\linewidth]{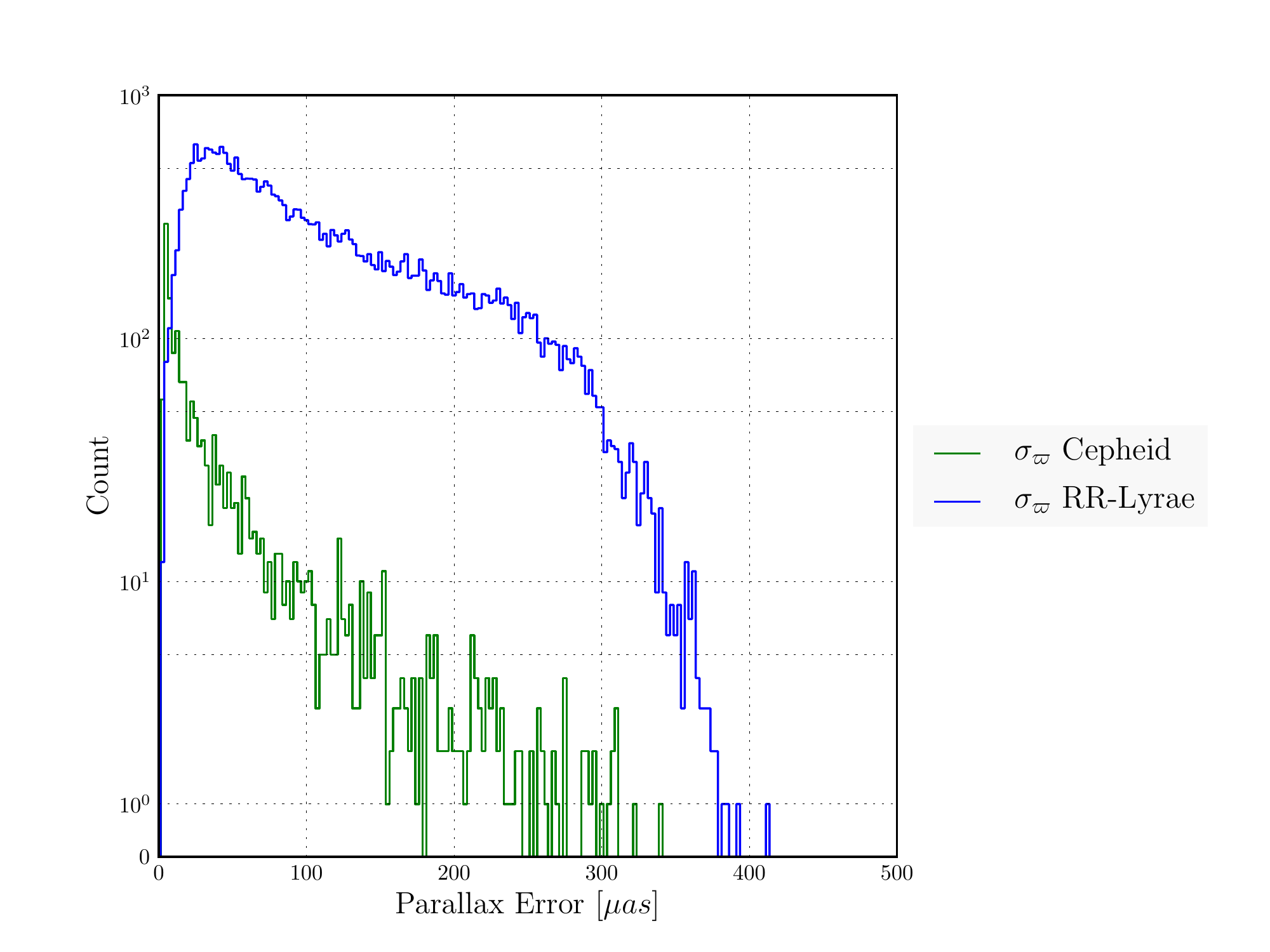} 
\caption{Histogram of parallax error for Cepheid and RR-Lyrae variable stars. RR-Lyrae is a combination of the two sub-populations RR-ab and RR-c.}
\label{fig:cepheidsPi}
\end{minipage}
\quad
\begin{minipage}[b]{0.45\linewidth}
\includegraphics[width=\linewidth]{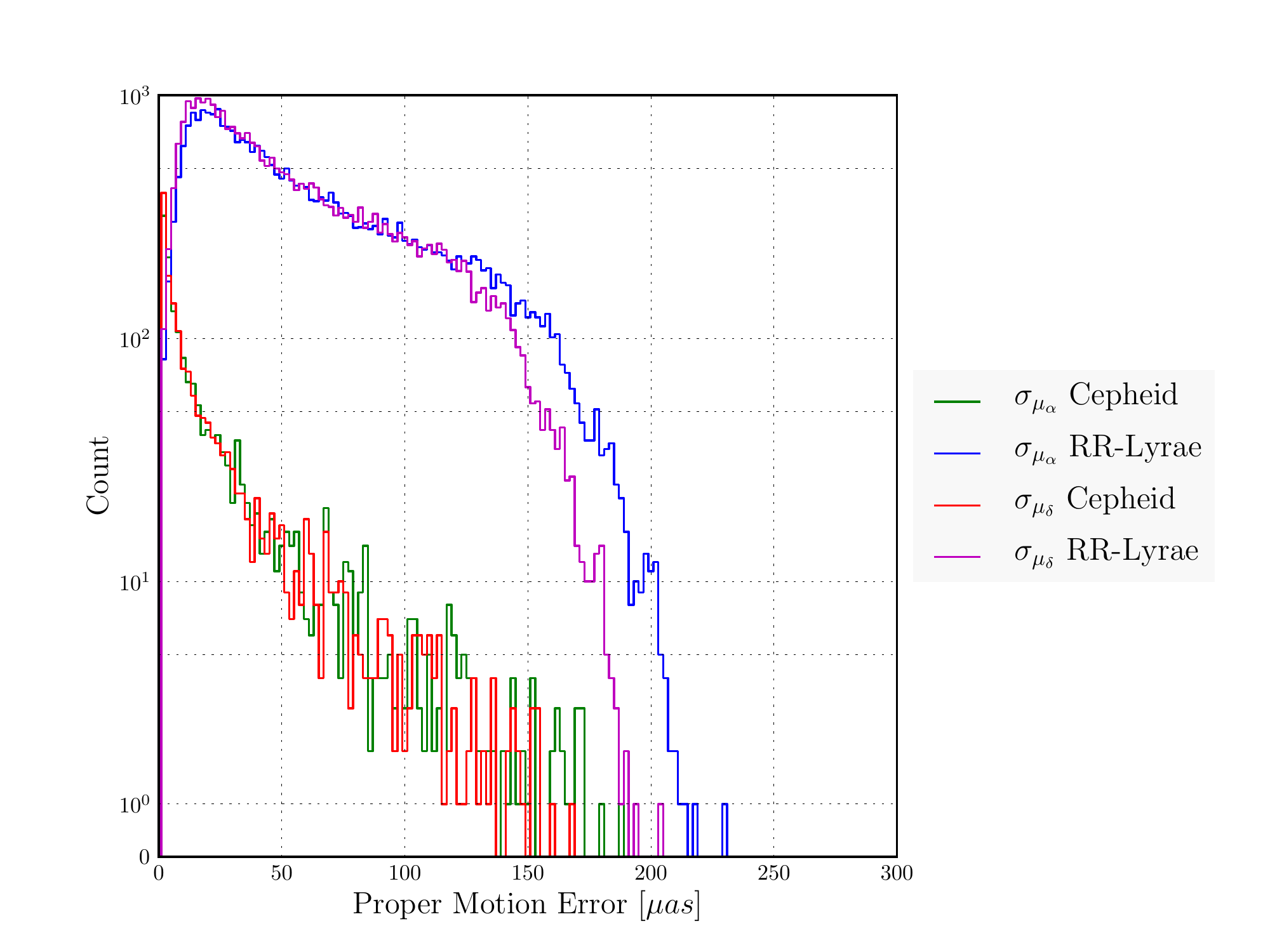} 
\caption{Histogram of proper motion error $\mu_\alpha$ and $\mu_\delta$ for Cepheid and RR-Lyrae variable stars. RR-Lyrae is a combination of the two sub-populations RR-ab and RR-c.}
\label{fig:cepheidsMu}
\end{minipage}
\end{figure*}

Cepheids and RR-Lyrae are types of pulsating variable stars. Their regular pulsation and a tight period-luminosity relation make them excellent standard candles, and therefore of particular interest in studies of galactic structure and the distance scale. Figure \ref{fig:cepheidsPi} shows the histogram of error in parallax specifically for Cepheid and RR-Lyrae variable stars, while Fig. \ref{fig:cepheidsMu} shows the errors in proper motions for Cepheids and RR-Lyrae.

\subsection{Physical parameters}\label{sec:Params}

Adding low-resolution spectral photometers on-board Gaia will make it capable of providing information on several object parameters including an estimate of line-of-sight extinction, effective temperature, metallicity, and surface gravity. Discussion of each individual physical parameter is given below.

Provided here are results for an approximation of the results of \cite{CU8}, which reproduces CU8 results statistically but not individually for each star. Therefore for detailed analysis of specific object types, care should be taken. Again, due to the very long tails of the error distributions caused by large numbers of extremely faint stars, the mean values given below should be taken with caution.

\begin{figure*}[ht]
\centering
\subfigure{%
\includegraphics[width=0.45\linewidth]{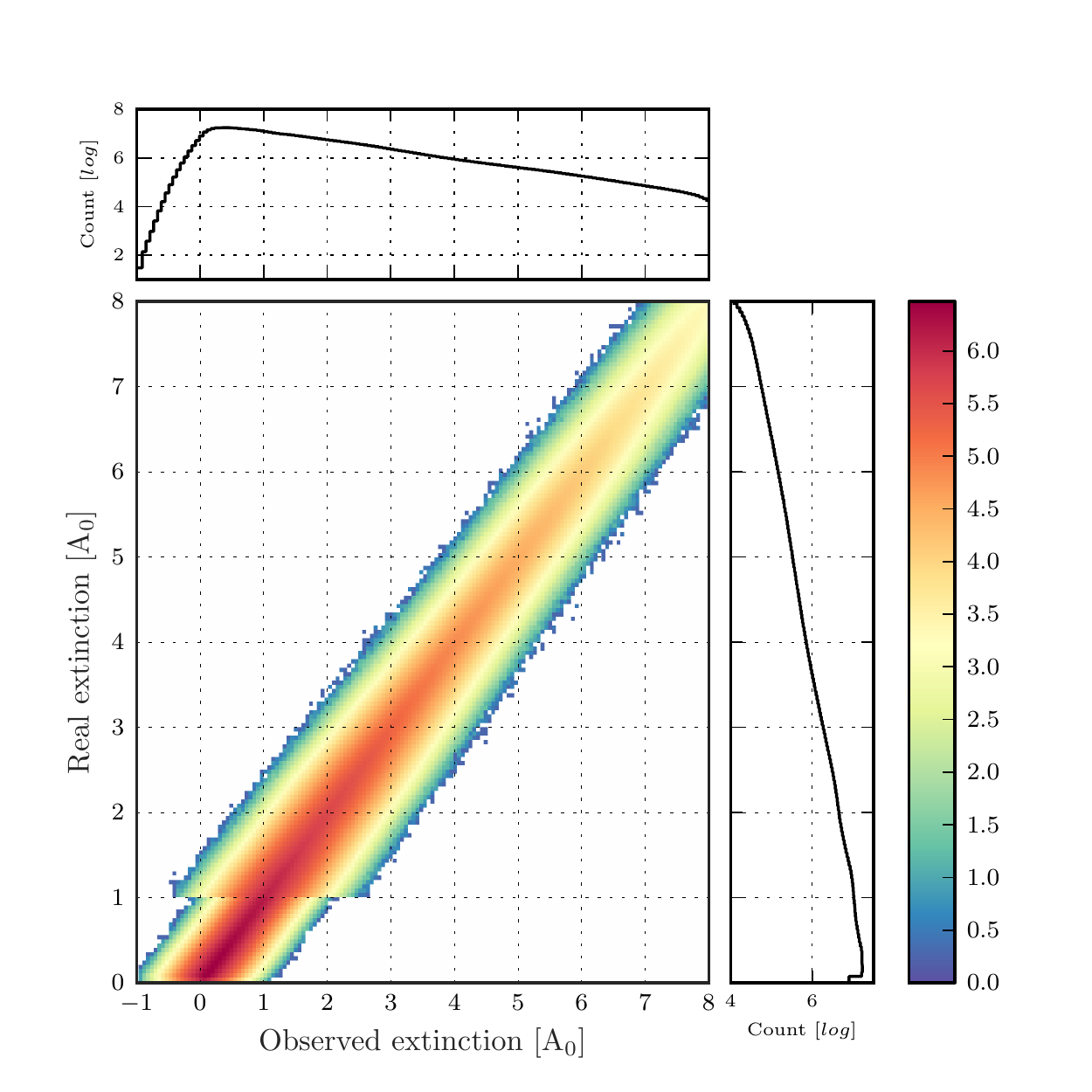} }
\quad
\subfigure{%
\includegraphics[width=0.45\linewidth]{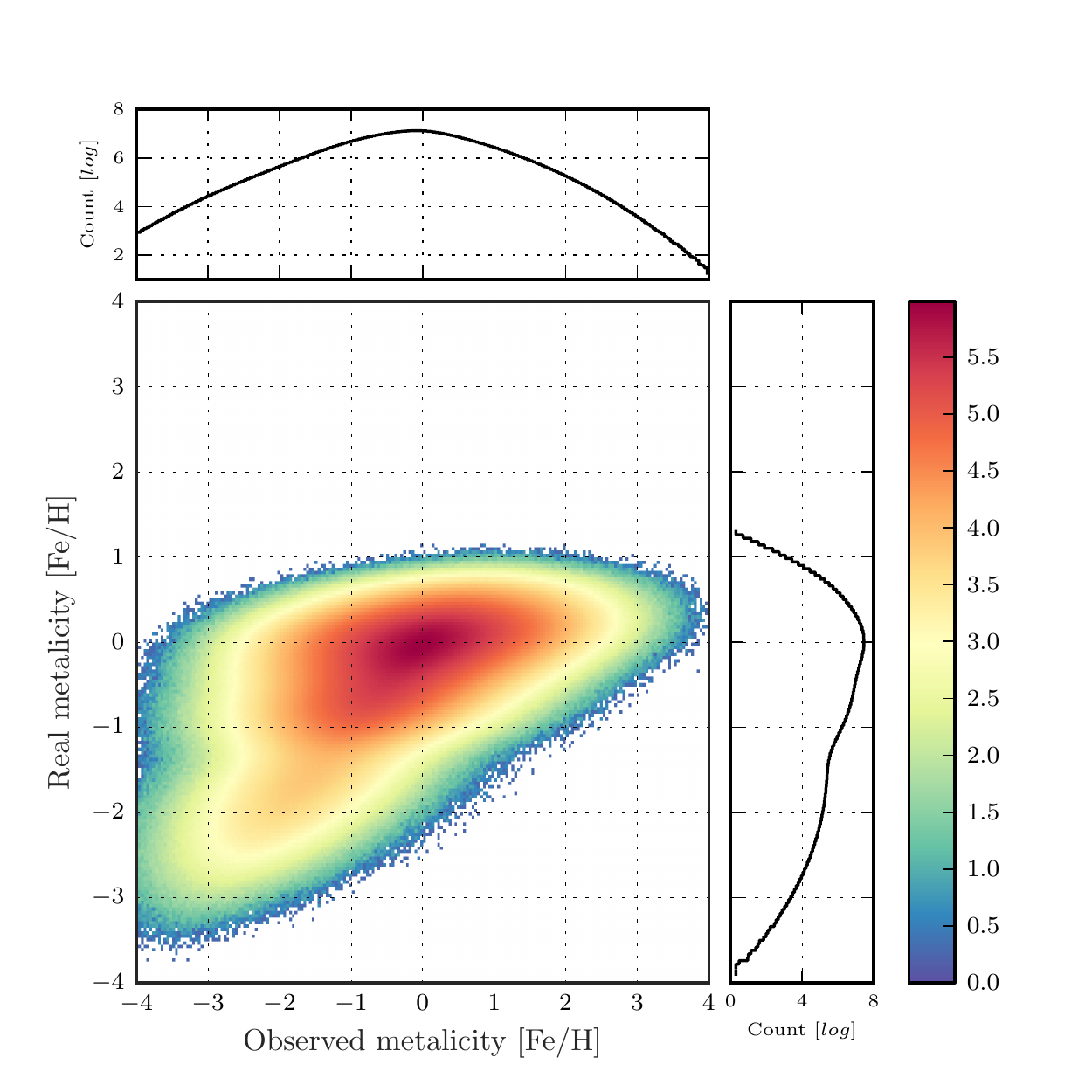} }
\subfigure{%
\includegraphics[width=0.45\linewidth]{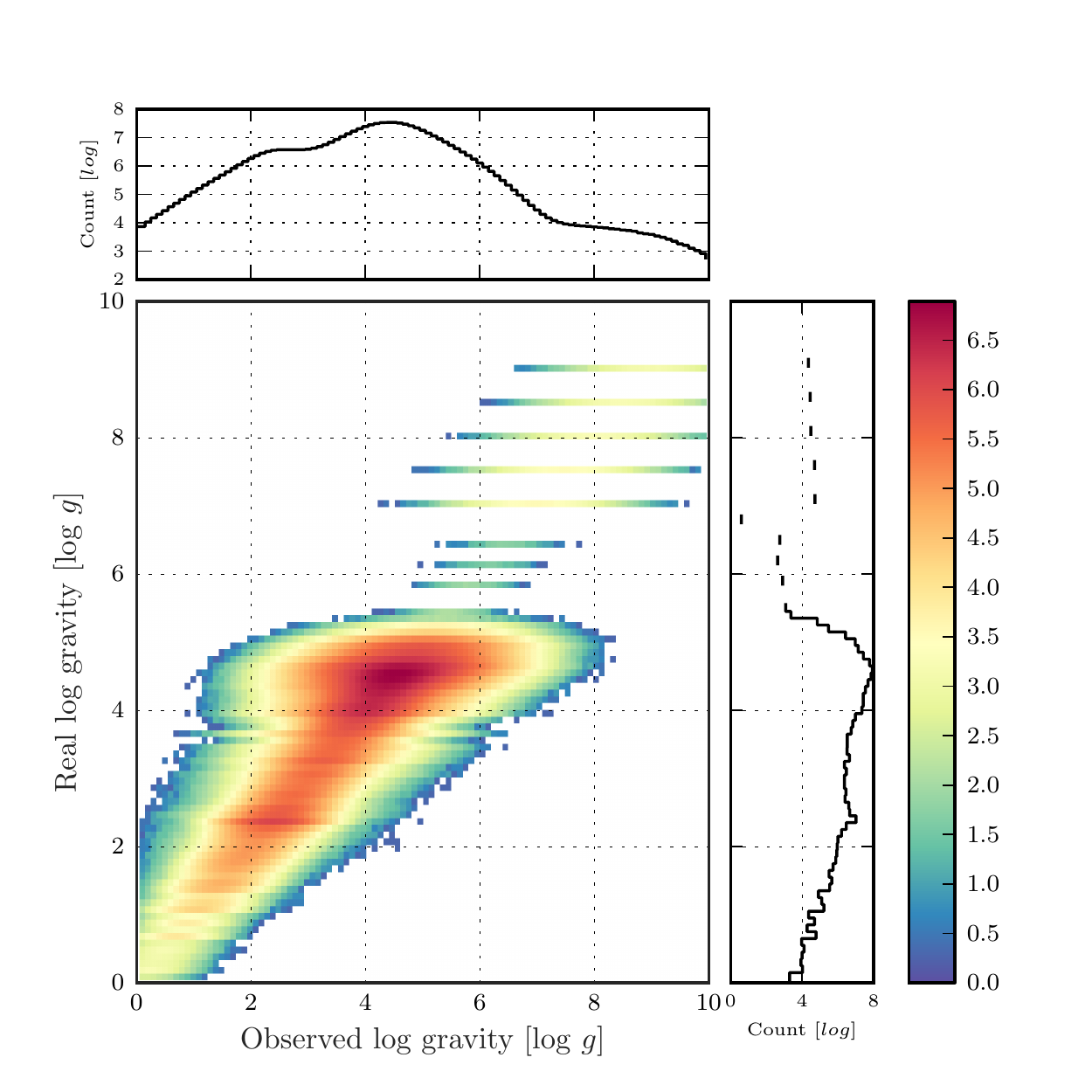} }
\quad
\subfigure{%
\includegraphics[width=0.45\linewidth]{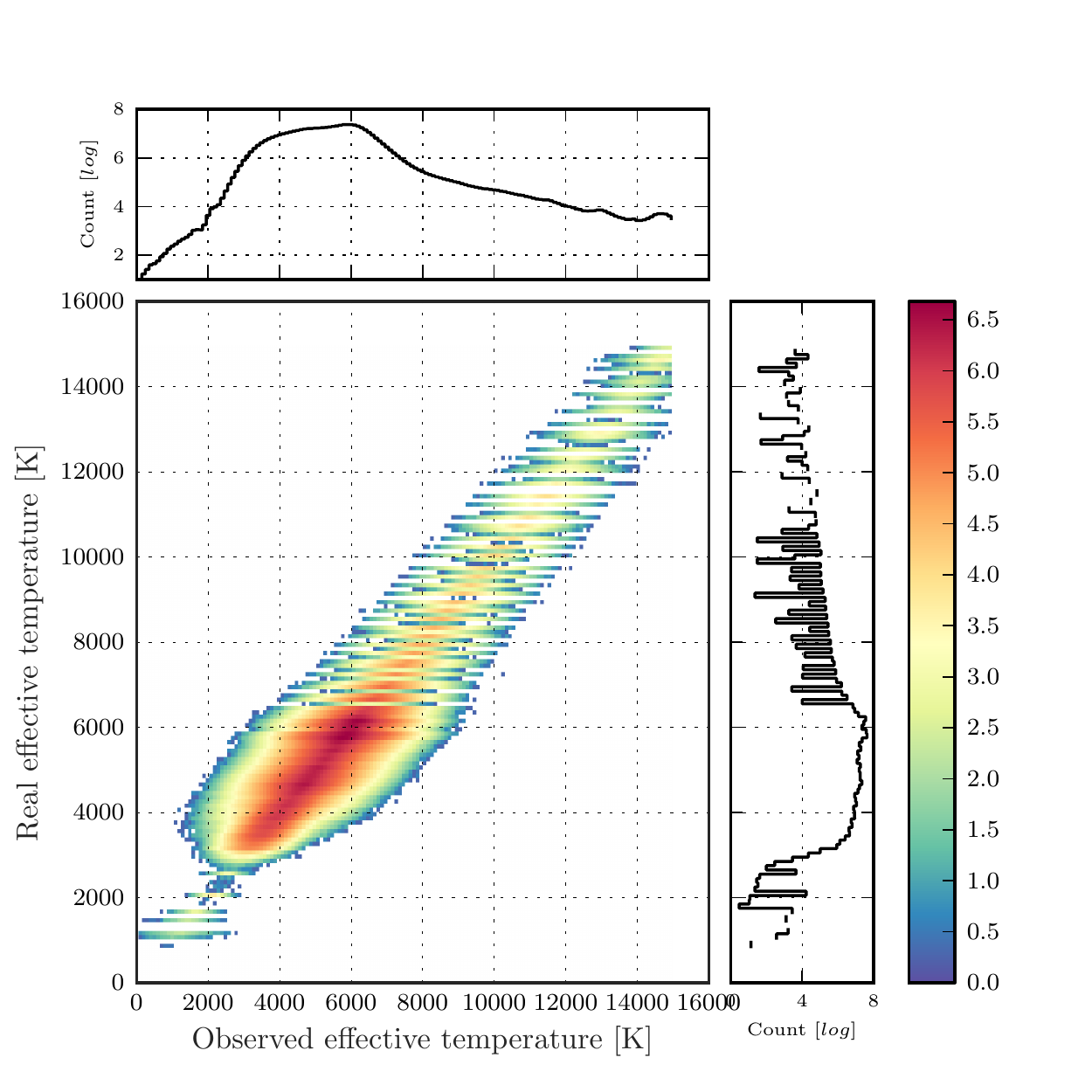} }

    \caption{Comparison of the true values of physical parameters with the GOG `observed' values for: \emph{Top left}, extinction; \emph{top right}, metallicity; \emph{bottom left}, surface gravity; and \emph{bottom right}, effective temperature. The colour scales represent log density of objects in a bin size of: \emph{top left}, 50 by 50 mmag; \emph{top right}, 0.4 by 0.4 dex; \emph{bottom left}, 0.5 by 0.5 dex; and \emph{bottom right}, 100 by 100 K. White area represents zero stars.}
    \label{fig:PhysicalParamsComparison}
\end{figure*}

\subsubsection{Extinction}

Across many fields of astronomy, the effects of extinction on the apparent magnitude and colour of stars can play a major role in contributing to uncertainty. An accurate estimation of extinction will prove highly useful for many applications of the Gaia Catalogue. 

Figure \ref{fig:PhysicalParamsComparison} shows the comparison between true extinction and the simulated Gaia estimate. For the vast majority of stars, the Gaia estimated extinction lies very close to the true value. This could prove very useful when, for example, using parallax and apparent magnitude data from Gaia because accurate extinction estimates are required to constrain the absolute magnitude of an object. 

Additionally, these results show that the Gaia data will be highly useful in terms of mapping galactic extinction in three dimensions, thanks to the combination of a large number of accurate parallax and extinction measurements. The negative extinction values in Fig. \ref{fig:PhysicalParamsComparison} are of course non-physical and are simply the result of applying a Gaussian random error to stars with near zero extinction.

The discontinuity at $A_0=1$ in the top left-hand panel of Fig. \ref{fig:PhysicalParamsComparison} comes from the distinction made between high and low extinction stars in the presentation of the results in \cite{CU8}. Our algorithm is based on results given in that paper, where the dependence on the extinction has been simplified to two cases, stars with $A_0<1$ and those with $A_0>1$. This distinction was made only for presentation of the results, and the real results from the DPAC algorithms will not show this discontinuity.
\cite{CU8} report a degeneracy between extinction and effective temperature due to the lack of resolved spectral lines sensitive only to effective temperature.

\subsubsection{Effective temperature}

For all objects in the GOG catalogue, the measured effective temperature ranges between 850 and 102000 K. The error in effective temperature is less than 640 K for all stars, with a mean value of 388 K. Figure \ref{fig:PhysicalParamsComparison} shows the comparison between true object effective temperature and the Gaia estimation. The thin lines visible in Fig. \ref{fig:PhysicalParamsComparison} are an artefact from the UM, which uses a Hess diagram to produce stars, leading to some quantisation in the effective temperature of simulated stars.

\subsubsection{Metallicity}

Metallicity can be estimated by Gaia in the form of [$Fe/H$]. Measured values range from -6.5 to +4.6. The mean error in metallicity estimate is 0.57 dex. The relatively high error in metallicity estimate can lead a large difference between real and observed values, as seen in Fig. \ref{fig:PhysicalParamsComparison}.

\subsubsection{Surface gravity}

The mean error in surface gravity is 0.45 dex. The comparison between real and observed surface gravity can be seen in Fig. \ref{fig:PhysicalParamsComparison}. As with metallicity, the lines at regular intervals at high gravity in this plot are due to the UM \citep{robin2011}.

\section{Conclusions}
\label{sec:conclusions}
The Gaia Object Generator provides the most complete picture to date of what can be expected from the Gaia astrometric mission. Its simulated catalogue provides useful insight into how various types of objects will be observed and how each of their observables will appear after including observational errors and instrument effects. The simulated catalogue includes directly observed quantities, such as sky position and parallax, as well as derived quantities, such as interstellar extinction and metallicity. 

Additionally, the full sky simulation described here is useful for gaining an idea of the size and format of the eventual Gaia Catalogue, for preparing tools and hardware for hosting and distribution of the data, and for becomeing familiar with working with such a large and rich dataset.

In addition to the stellar simulation described in this paper, there are plans to generate other simulated catalogues of interest, such as open clusters, Magallanic Clouds, supernovae, and other types of extragalactic objects, so that a more complete version of the simulated Gaia Catalogue can be compiled.

Here we have focussed on the simulated catalogue from the inbuilt Gaia Universe Model, based on the Besan\c{c}on Galaxy model. However, GOG can alternatively be supplied with an input catalogue generated by the user. This way, simulated data from any other model can be processed with GOG to obtain simulated Gaia observations of specific interest to the individual user. The input can be either synthetic data on a specific star or catalogue, or an entire simulated survey such as those generated using Galaxia \citep{Galaxia}, provided a minimum of input information is supplied (e.g. position, distance, apparent magnitude, and colour).

With GOG, the capabilities of the instrument can be explored, and it is possible to gain insight into the expected performance for specific types of objects. While only a subset of the available statistics have been reproduced here, it is possible to obtain the full set of available statistics at request. 

We are working to make the full simulated catalogue publicly available, so that interested individuals can begin working with data similar to the forthcoming Gaia Catalogue.

\begin{acknowledgements}

GOG is the product of many years of work from a number of people involved in DPAC and specifically CU2. The authors would like the thank the various CUs for contributing predicted error models for Gaia. The processing of the GOG data made significant use of the Barcelona Supercomputing Center (MareNostrum), and the authors would specifically like to thank Javi Casta\~{n}eda, Marcial Clotet, and Aidan Fries for handling our computation and data handling needs. Additionally, thanks go to Sergi Blanco-Cuaresmo for his help with Matplotlib.

This work was supported by the Marie Curie Initial Training Networks grant number PITN-GA-2010-264895 ITN ``Gaia Research for European Astronomy Training'', and MINECO (Spanish Ministry of Economy) - FEDER through grant AYA2009-14648-C02-01, AYA2010-12176-E, AYA2012-39551-C02-01, and CONSOLIDER CSD2007-00050. 

\end{acknowledgements}

\bibliographystyle{aa} 
\bibliography{References} 

\end{document}